\documentclass[12pt,letterpaper]{article}

\usepackage[square,authoryear]{natbib}
\usepackage[pdftex,dvipsnames]{xcolor}  % Coloured text etc.
\usepackage{setspace}
\usepackage{graphicx}
\usepackage{subfig}
\usepackage{float}
\usepackage{background}
\usepackage{xargs}                      % Use more than one optional parameter in a new commands
\usepackage{hyperref}

\usepackage[colorinlistoftodos,prependcaption,textsize=tiny]{todonotes}
\newcommandx{\unsure}[2][1=]{\todo[linecolor=red,backgroundcolor=red!25,bordercolor=red,#1]{#2}}
\newcommandx{\change}[2][1=]{\todo[linecolor=blue,backgroundcolor=blue!25,bordercolor=blue,#1]{#2}}
\newcommandx{\info}[2][1=]{\todo[linecolor=OliveGreen,backgroundcolor=OliveGreen!25,bordercolor=OliveGreen,#1]{#2}}
\newcommandx{\improvement}[2][1=]{\todo[linecolor=Plum,backgroundcolor=Plum!25,bordercolor=Plum,#1]{#2}}
\newcommandx{\ensure}[2][1=]{\todo[linecolor=YellowOrange,backgroundcolor=YellowOrange!25,bordercolor=YellowOrange,#1]{#2}}

\backgroundsetup{contents=Work-in-progress,opacity=0.3,scale=5}

\begin{document}

\title{Foundations and Scoping of Data Science\thanks{This is work-in-progress as the various notes in the text indicate. I have decided to release this early version, because of the interest in the topic and because I realized that it will take me a while to complete it. I will release new versions as I reach meaningful milestones. Feedback most welcome. \\A shorter version of this has appeared as \citep{Ozsu:2023aa}; please cite that paper when appropriate.}}

\author{M. Tamer \"{O}zsu \\
\small University of Waterloo, Canada \\
\href{mailto:tamer.ozsu@uwaterloo.ca}{\small \texttt{tamer.ozsu@uwaterloo.ca}}}
%tamer.ozsu{\fontfamily{ptm}\selectfont @}uwaterloo.ca}
\date{\today}
%%
%% This command processes the author and affiliation and title
%% information and builds the first part of the formatted document.
\maketitle

%\linenumbers

\begin{abstract}
There has been an increasing recognition of the value of data and of data-based decision making. As a consequence, the development of data science as a field of study has intensified in recent years. However, there is no systematic and comprehensive treatment and understanding of data science. This article describes a systematic and end-to-end framing of the field based on an inclusive definition. It identifies the core components making up the data science ecosystem, presents its lifecycle modeling the development process, and argues its interdisciplinarity. 

(I will write a better abstract in due course.)
\end{abstract}

\section{Introduction}
\label{sec:intro}

There is a data-driven revolution underway in science and society, disrupting every form of enterprise. We are collecting and storing data more rapidly than ever before. The value of data as a central asset in an organization is now well-established and generally accepted. Economist called it ``the world's most valuable resource''~\citep{Economist2017}. %\footnote{\textit{Economist} Cover Story, 6 May 2017 Issue.} 
World Economic Forum's briefing paper \emph{A New Paradigm for Business of Data} states ``At the heart of digital economy and society is the explosion of insight, intelligence and information -- data.'' ~\citep{Forum:2020vo} Businesses have also been commenting on the  value of data to their operations: ``data is the new oil'' (Clive Robert Humby, Chief Data Scientist, Starcount), ``data is the new currency'' (Antonio Neir, President, Hewlett-Packard Enterprise), ``data is a commodity like gold'' (Matt Shephard, Head of Data Strategy, BBH London). 

The field of \textit{data science} is expected to enable data to be leveraged for making better  decisions and achieving more meaningful outcomes. However, full exploitation of the potential has been limited by the lack of knowledge exchange between experts in the sub-fields of data science, and by the lack of tools, methods, and principles for understanding and translating these insights into improved decisions, products, systems, and policies. Although the term data science has some history (see Section \ref{sec:ds-def}), in its current incarnation as a modern field of study, it has already had significant economic impact. McKinsey Global Institute estimates that the field of data science is currently generating \$1.3 trillion in economic value every year in the US; other countries have reported proportionally similar contributions. A 2015 Organisation for Economic Co-operation and Development (OECD) report identified ``data-driven innovation'' (DDI) as having a central driving role in twenty-first century economies, defining DDI as ``the \textbf{use of data and analytics} to improve and foster new products, processes, organisational methods and markets'' [Emphasis from original report]. Data science deployments are still what might be called first-generation, but their impact in many areas are already being felt: global sustainability~\citep{Dunn:2021wh}, power and energy systems~\citep{Bangert:2021um}, biological and biomedical systems~\citep{Supriya2021}, health sciences and health informatics~\citep{Consoli:2019uh}, finance and insurance~\citep{Chakravaram:2019tg}, smart cities~\citep{Sarker:2022ud}, digital humanities~\citep{Milligan:2019tu} and others (see Section \ref{sec:apps}).

Increasing number of countries have released policy statements related to data science. US has a number of initiatives: White House Open Data Initiative~\citep{WH2013}, Department of State Data Policy~\citep{USState2021}, and National Artificial Intelligence Task Force~\citep{WH2021} are examples. European Union's data strategy~\citep{EU2022} includes a number of initiatives around data science. Canada has released a federal digital charter~\citep{CA2020} that addresses the importance of data in the Canadian economy and establishes a vision for a data economy built around ten principles (universal access; safety and security; control and consent; transparency, portability, and interoperability; open and modern digital government; a level playing field; data and digital for good; strong democracy; free from hate and violent extremism; and strong enforcement and real accountability). The importance of this charter is in its multidisciplinarity and the need for engagement by a number of academic disciplines. 

The recent COVID-19 pandemic has clarified and emphasized the importance of data-based decision-making. Individual countries have differed in how well they have gathered, analyzed, and used data to make public health decisions; but globally, we have not done well, and this should be a major research agenda for the data science field~\citep{Chakravorti:2022aa}.

\begin{figure}
\centering
\includegraphics[width=0.9\linewidth]{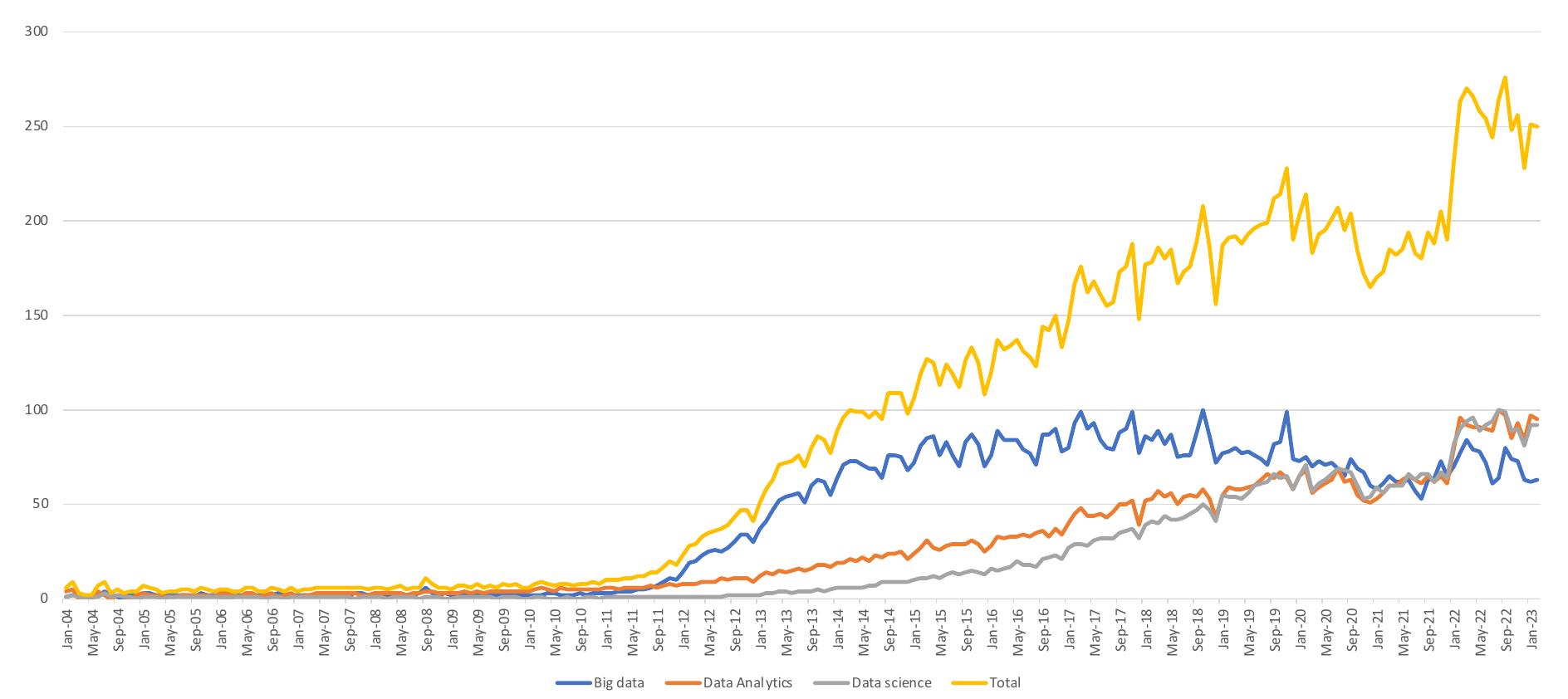}
\caption{Trending of data science-relevant terms\\Source: Google Trends on January 2023}
\label{fig:trend}
\end{figure}

The last decade has established ``big data'' and ``data science'' into our lexicon, both as buzzwords and as important areas of study. Interest in the topic, as evidenced by Google Trends (Figure \ref{fig:trend}), has exploded over the same time period.  An increasing number of countries have released policy documents related to data science. In academia, data science programs and research institutes have been established with significant speed, while many industrial organizations have created data science units. A quick survey of these programs and initiatives suggests a common core, but also a lack of unified framing of data science. Many of the programs and initiatives find cleavages in their respective organizations and shape themselves to fit these. While this may ease organizational acceptance, and is, to a certain extent, unavoidable, we should be operating with more clarity. %At a recent Data Science Leadership Summit  (which is run by the Academic Data Science Alliance), a colleague opined that no one should try to form a data science institute or centre without first defining what ``data science'' means in the context of their organization. It is hard to disagree with that sentiment. The purpose of this article is to tackle this issue and provide a comprehensive framework for data science as a field.

There are a number of reasons why this clarification is helpful. An important one is to be able to understand whether data science is an academic discipline. This is hard to know without a  definition of data science and identification of its core and scope. A related reason is to provide an intellectually consistent framing to the numerous data science institutes and academic units being formed. A third reason is to bring some clarity to the question of who a data scientist is. The point is not to constrain what is meant by a data scientist or to limit the scope of these academic initiatives, but to acknowledge the diversity around some commonalities. A fourth and final reason is that  a systematic investigation of the field is likely to identify important techniques and tools that should be in a professional data scientist's toolbox. 

\begin{figure}
\centering
\includegraphics[width=0.8\linewidth]{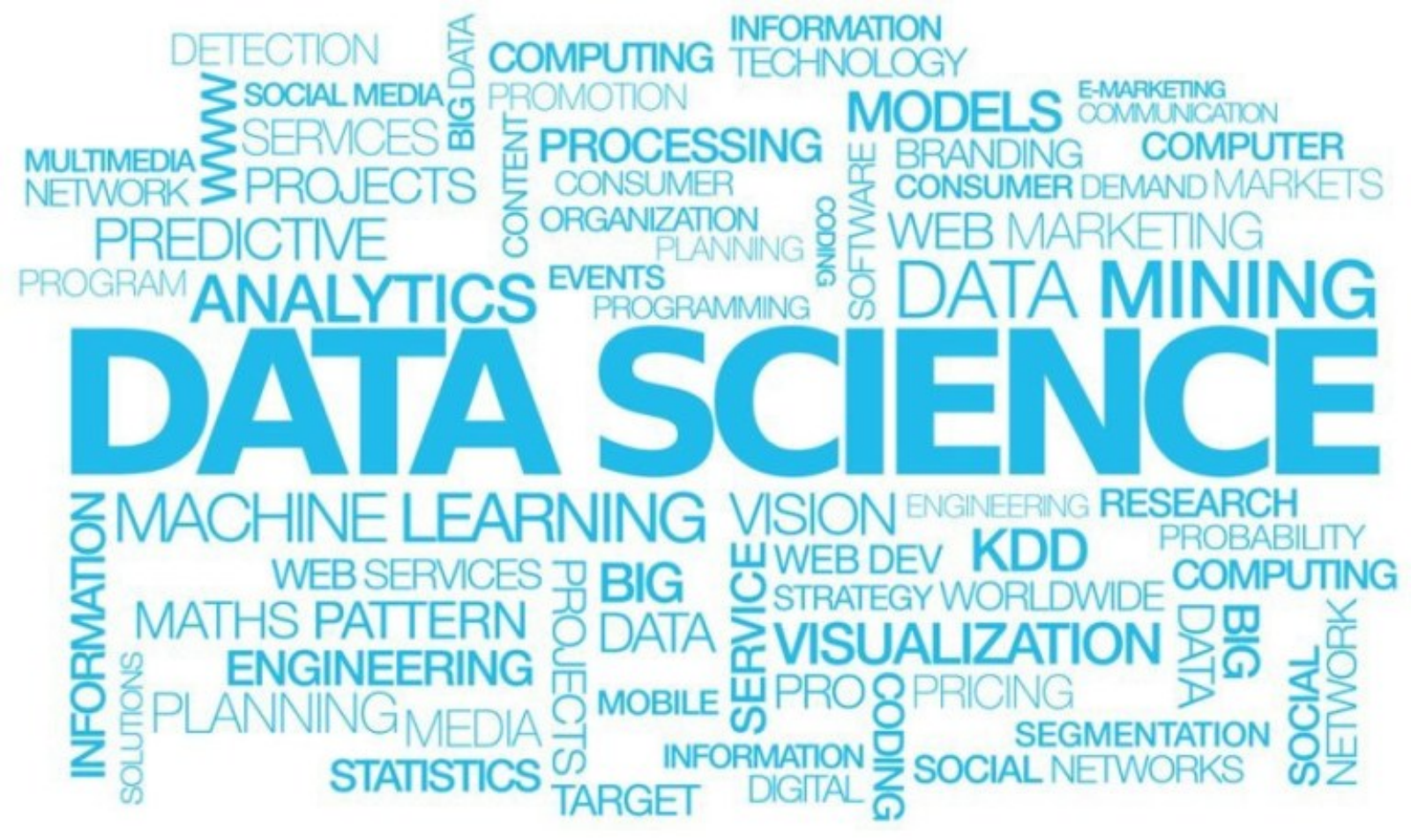}
\caption{Data Science Word Cloud}
\label{fig:dscloud}
\end{figure}

Part of the difficulty is the carelessly interchangeable use of the terms ``big data,'' ``data analytics,'' and ``data science'' in much of the popular literature, which frequently spills over to technical literature, as evidenced by the word cloud that was generated a few years ago (Figure \ref{fig:dscloud}). It is important to get them right. Data analytics, as defined in the next two sections, is one component of data science and not synonymous with it. Data science is not the same as big data. Perhaps the best analogy between them is that big data is like raw material; it has considerable promise and potential if one knows what to do with it. Data science gives it purpose, specifying how to process it to extract its full potential and to what end. It does this typically in an application-driven manner, allowing applications to frame the study objective and question. Applications are central to data science; if there are no applications to drive the inquiry, it is hard to argue that there is a data science deployment. Jagadish also emphasizes this point, stating `` `Big Data' begins with the data characteristics (and works up from there), whereas `Data Science' begins with data use (and works down from there).''~\citep{Jagadish:2015ul}

A second difficulty is the vagueness in many definitions about the relationship between data science, machine learning (ML), and data mining (DM) arising from the colloquial use of ``data science'' to mean data analytics using ML/DM. Data science is not a subfield of  ML/DM nor is it synonymous with these disciplines. I return to this topic in some detail in the following section.
%More broadly, it is not the case that data science is a subtopic of AI -- a common claim originating from confusion on boundaries. AI and data science are conceptually different fields that overlap when ML/DM techniques are used in data analytics,  but otherwise have their own broader concerns. The broader scope of data science is discussed in this paper, highlighting its constituents that are not part of AI. Conversely, there are topics in AI, such as agents, robotics, automated programming and others, that are not within the scope of data science. Thus, AI and data science are related, but one does not encompass the other.

A final difficulty is that data science is broad in scope, involving a number of disciplines and finding the right synthesis is not straightforward. At the risk of oversimplification, the following are the different constituencies that have an interest in data science: (1) STEM people who focus on foundational techniques and the underlying principles (computer scientists, mathematicians, statisticians), (2) STEM people who focus on science and engineering data science applications (e.g., biologists, ecologists, earth and environmental scientists, health scientists, engineers), and (3) non-STEM people who focus on social, political, and societal aspects. It is important to include all these constituencies in the discussions surrounding data science, while establishing a recognizable core of the field. This is a difficult balance to maintain. 

My objective in this article is to put forth an internally consistent and coherent view of data science. The article discusses core components,  contributing disciplines,  lifecycle considerations, and  shareholder communities. The main takeaways can be summarized as follows:

\begin{enumerate}
\item It is important to clearly establish a consistent and inclusive view the entire field, and one is proposed.
\item To avoid becoming a catch-all or whatever the particular circumstances allow, it is essential to define the core of the field while being inclusive, and four core areas are identified.
\item It is critical to  take a holistic view of activities that comprise data science.
\item A framework must be established that facilitates cooperation and collaboration among a number of disciplines. 
\end{enumerate}

Data science is still in its early days as an emerging field. This article contributes to the discussions around its nature and scope. There will, hopefully, be joinders to the discussion to better define the field.

\section{What is Data Science?}
\label{sec:ds-def}

The origins of the term \textit{data science} are fuzzy. Data is central to both statistics and computing, so both communities have attempted to define the field.

Statisticians suggest that its origins lead back to John Tukey~\citep{Tukey:1962ue} who passionately argued for the separation of ``data analysis'' from ``classical statistics''. His main point was that data analysis is an empirical science while classical statistics is pure mathematics. Tukey defines data analysis as ``procedures for analyzing data, techniques for interpreting the results of such procedures, ways of planning the gathering of data to make its analysis easier, more precise or more accurate, and all the machinery and results of (mathematical) statistics which apply to analyzing data.'' Tukey's focus is on the analytics component of data science, and despite the breadth of this definition, it does not capture the fullness of data science as argued in this paper. The important part of Tukey's viewpoint, however, is his acknowledgement of the process aspect of data analysis, which is also important for the broader data science definition.

The term ``data science''  emerges somewhat later in statistics literature in a paper by Chikio Hayashi, who states: ``data science intends to analyze and understand actual phenomena with `data'. In other words, the aim of data science is to reveal the features or hidden structure of complicated natural, human, and social phenomena with data from a different point of view from the established or traditional theory and method'' \cite{Hayashi:1998aa}. Hayashi continues Tukey's distinction of data science from ``traditional [statistical] theory and method.'' The roots of what we now consider a fundamental characteristic of data science, namely that data and its analysis are at the core of understanding actual phenomena, are clearly stated.

Capturing a precise definition of data has been important from the start of computing as a discipline. IFIP (International Federation of Information Processing) definition of data is ``a representation of facts or ideas in a formalized manner capable of being communicated or manipulated by some process.''~\citep{Gould:1971tn}. Naur builds his definition on this: ``data science is the science of dealing with data, once they have been established, while the relation of data to what they represent is delegated to other fields and sciences.'' ~\citep{Naur:1974uc}. 

Clearly, both statisticians and computer scientists have been thinking of data science for a long time and the understanding of what it is has evolved over time. A subtle shift happened in 2000s with the recognition that data science is broader than data analytics and that it involves a process from data ingestion to the production of insights and actionable recommendations. During this period, data-intensive approaches to problem solving have started to produce results. This became known as the ``big data revolution'' and resulted in data-centric approaches in many fieleds: natural language processing (e.g., \citep{Banko:2001tt} is an early work), computer vision (e.g., \citep{Liu:2016wn}), computational astronomy (e.g., \citep{Szalay:2002tm}), computational astrophysics (e.g., \citep{Naab:2017wu}), computational biology (e.g., \citep{Kitano:2002tz}), and many others. What initially began as a computational paradigm (also called the third paradigm), where computational methods could replace or enhance laboratory experimental methods\footnote{A 2001 New York Times article declared ``all science is computer science''~\citep{Johnson:2001uu}.}, rather quickly changed to data-intensive methods. This is frequently referred as the fourth paradigm~\citep{hey2009the}, and data science systematizes this understanding. Gray has referred to this alternatively as ``eScience'' ~\citep{Gray:2007aa}.

There are significant differences between what was called data analysis (or analytics) and what the current understanding of data science entails. More modern definitions of data science encompass this broader interpretation -- for example: ``Data science encompasses a set of principles, problem definitions, algorithms, and processes for extracting non-obvious and useful patterns from large datasets.''~\citep{Kelleher:2018tv} \cite{Cao:2017aa} data science as ``science of data'' or ``the study of data.'' He complements this general definition with a process view: ``data science is a systematic approach 
to `thinking with wisdom,' `understanding domain,' `managing data,' `computing with data,' `mining on knowledge,' `communicating with stakeholders,' `delivering products,' and `acting on insights'.'' \citep{Cao:2016aa}. As discussed below, it is important to capture the process view of data science.

%also provides a comprehensive definition that highlights this difference: ``\ldots dat ascience is a new interdisciplinary field that synthesizes and builds on statistics, informatics, computing, communication, amanagement and sociology to study data and its environments (including domains and other contextual aspects, such as organizational and social aspects) in order to transform data to insights and decisions by allowing a data-to-knwoledge-to-wisdom thinking and methodology.''

The The National Consortium for Data Science (NCDS), which is a U.S. consortium of leaders from academia, industry, and government, defines data science as the ``systematic study of organization and use of digital data for research discoveries, decision-making, and data-driven economy''~\citep{Ahalt:2013wr}. ACM (Association for Computing Machinery), in its proposal for an undergraduate data science degree, does not attempt to define data science directly, but indicates that it requires ``a strong focus on data -- the
collection of data and, through analysing it appropriately, using this to bring about beneficial insights and changes.''~\citep{Force:2021aa}.

The popularity of the current wave of data science approaches in scientific endeavors are matched by industrial interest with many companies forming data science teams and laying out their vision of data science in their organizations. It is instructive to briefly consider some of these. Note, in particular, the emphasis on multi-disciplinarity and breadth.

IBM:  ``Data science is a multidisciplinary approach to extracting actionable insights from the large and ever-increasing volumes of data collected and created by today’s organizations. Data science encompasses preparing data for analysis and processing, performing advanced data analysis, and presenting the results to reveal patterns and enable stakeholders to draw informed conclusions.''~\citep{IBMEducation:tv}

Oracle: ``Data science combines multiple fields, including statistics, scientific methods, artificial intelligence (AI), and data analysis, to extract value from data. … Data science encompasses preparing data for analysis, including cleansing, aggregating, and manipulating the data to perform advanced data analysis.''~\citep{Oracle:wu}

DataRobot: ``Data science is the field of study that combines domain expertise, programming skills, and knowledge of mathematics and statistics to extract meaningful insights from data. \ldots In turn, these systems generate insights which analysts and business users can translate into tangible business value.''~\citep{DataRobot:tv}

Clearly, some common themes are emerging that would help identify the field. There is, however, also uncertainty in setting boundaries for the field. Kelleher and Tierney state ``the terms data science, machine learning, and data mining are often used interchangebly''~\citep{Kelleher:2018tv}. Wikipedia definition starts well: ``data science, also known as data-driven science, is an interdisciplinary field of scientific methods, processes, algorithms, and systems to extract knowledge or insights from data in various forms, either structured or unstructured'', captures a number of important characteristics, but then states ``similar to data mining.'' Leaving aside the fact that ``similar'' is an entirely unhelpful phrase in this context, roughly equating data science to data mining is very limiting. In some cases, the definitions are too vague and broad to be useful in identifying the core of the field: ``data science is an umbrella term to describe the entire complex and multistep processes used to extract value from data''~\citep{Irizarry:2020uw}. This is consistent with Irizarry's view that data science is not a discipline, but an umbrella term. Even if one accepts that data science is not (yet) a discipline, there is a long separation between a discipline and an ``umbrella term.'' 

In some cases, people have chosen to frame the field by describing what it is not: ``Data science is not all about using data for prediction or merely about data analysis. It is not a discipline confined to science, technology, engineering, and mathematics field, and, most imortantly, it is not a single discipline at all''~\citep{Ahmad:2022tt}. There is a lot that is correct in this description of what the field is not.

It has been argued that it is difficult to define data science, because it is multifaceted: it can be viewed as a science, as a research paradigm, as a research method, as a discipline, or a workflow~\citep{Mike:2023aa}. The multifaceted nature of data science is real. However, it is hard to make progress as a field (much less an academic discipline) unless the core is properly defined and the scope is accurately identified.

It is possible to be more precise, capturing both the core of the field and its breadth. The definitions discussed so far capture a number of different aspects of the field:

\begin{itemize}
\item interdisciplinarity (as defined in ~\citep{Choi:2006aa,Alvargonzalez:2011wm}),
\item a data-based approach to problem solving,
\item the use of large and multi-modal data,
\item the focus on deriving insights and value by discovering patterns and relationships in data, and
\item the underlying process-oriented lifecycle
\end{itemize}

A working comprehensive definition that captures the essence of the field and explicitly recognizes that it involves a process would be:

\begin{quote}
Data science is a data-based approach to  problem solving by analyzing and exploring large volumes of possibly multi-modal data, extracting knowledge and insight from it, and using information for better decision-making. It involves the process of collecting, preparing, managing, analyzing, and explaining the data and analysis results.
\end{quote}

The extraction of knowledge and insight for better decision making has also been called generating ``data products''~\citep{Loukides:2010wu}. This comprehensive definition captures both the essence of the field and explicitly recognizes that it involves a process. It is consistent with the current understanding of the broad scope of the field (see, for example, CRA's use of the term ~\citep{Getoor:2016ub}). 

Note that the definition intentionally uses the term ``data-based'' rather than the more common ``data-driven.'' The latter has frequently been interpreted as ``data should be the main (only?) basis for decisions'' since ``data speaks for itself.'' This is wrong -- data certainly holds facts and can reveal a story, but it only speaks through those who interpret it, and that can introduce biases. Therefore, data should be one of the inputs to decision making, not the only one. Furthermore, ``data-driven'' has come to mean that it is possible to take data and analyze it by automated tools to generate automated actions. This also is problematic despite the recent popularity of, and over-reliance on, predictive and prescriptive analytics. Though data science has significant potential, and successful data science application are plentiful (see Section \ref{sec:apps}), there are sufficient misuses of data to give us pause and concern -- Google's algorithmic detection of influenza spread using social media data is one prominent example, the Risk Needs Assessment Test used in U.S. justice system is another. Therefore, ``data-based'' is a preferable phrase that signals that these are aids to the decision-maker, not decision-makers themselves\footnote{Other suitable terms that have been used are ``data-enhanced'' or ``data-enabled''.}.

The definition also intentionally excludes  a number of misconceptions that pepper daily discourse and sometimes generate considerable debate. These may be called the ``myths of data science'' and it is important to highlight why the definition excludes or makes no reference to them.

First, is the point I raised in the previous section, namely that data science is not the same as (or synonymous with) data mining (DM) or machine learning (ML). These are techniques that are useful in data analytics, but data science is broader than data analytics (more on this in the next section). To expand this statement, it is not the case that data science is a subtopic of AI -- a common statement originating from confusion on boundaries. AI and data science are conceptually different fields that overlap when DM/ML techniques are used in data analytics,  but otherwise have their own broader concerns. Thus, they are related, but one does not encompass the other.

Second, data science is not the same as big data. Perhaps the best analogy between them is that big data is like raw material; it has considerable promise and potential if one knows what to do with it. Data science gives it purpose, specifying how to process it to extract its full potential and to what end. It does this typically in an application-driven manner, allowing applications to frame the study objective and question. Applications are central to data science; if there are no applications to drive the inquiry, it is hard to argue that there is a data science application/deployment. Jagadish also emphasizes this point and states `` `Big Data' begins with the data characteristics (and works up from there), whereas `Data Science' begins with data use (and works down from there).''~\citep{Jagadish:2015ul}

\section{Data Science Ecosystem}
\label{sec:eco}

\begin{figure}[ht]
\centering
\includegraphics[width=0.9\linewidth]{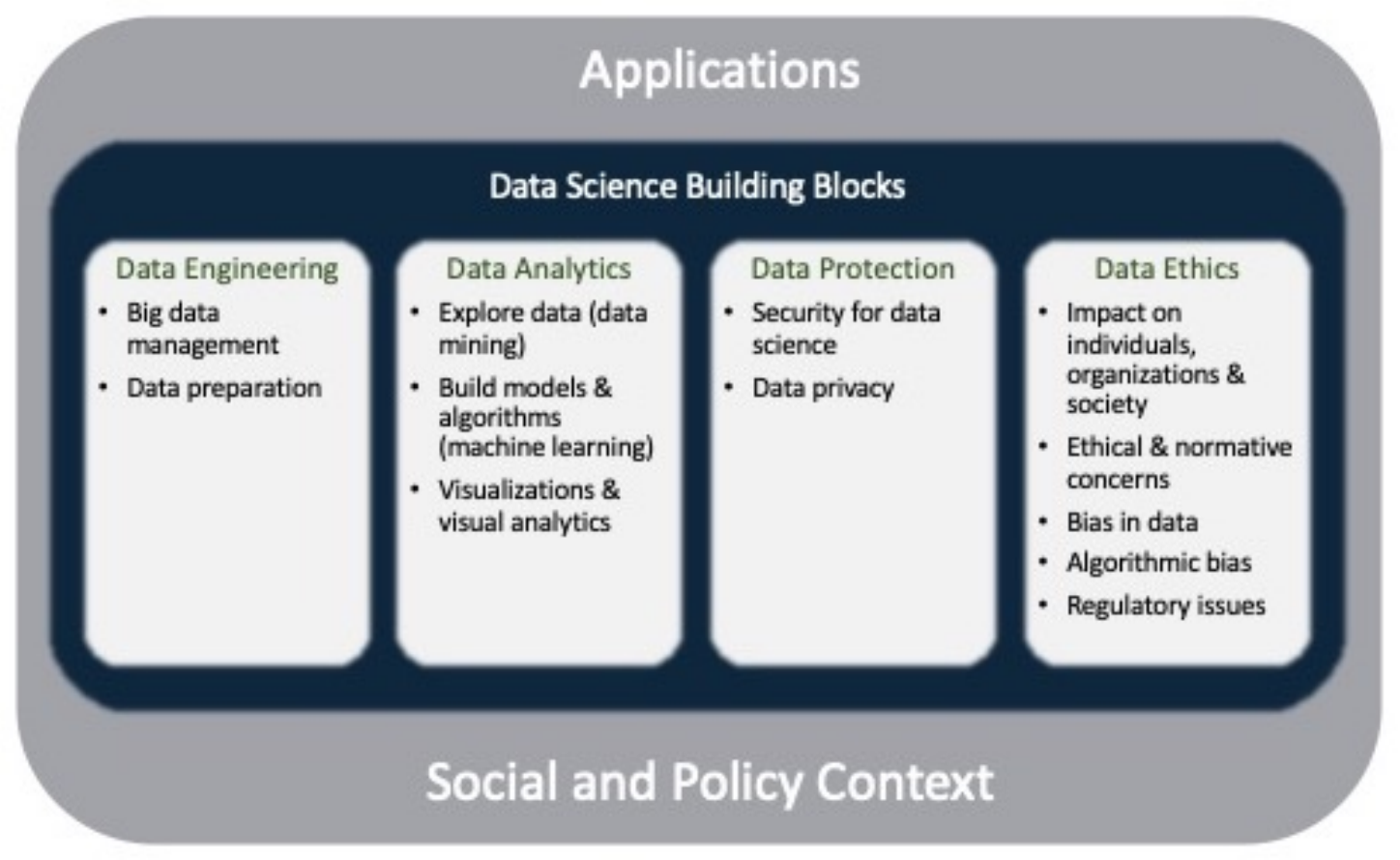}
\caption{Data Science Building Blocks.}
\label{fig:building}
\end{figure}

Data science is inherently interdisciplinary: it builds on a core set of capabilities in data engineering, data analytics, data protection, and data ethics (Figure \ref{fig:building}); these are the four pillars of data science. Some of this core is technical, some is not. Although the term ``data science'' is frequently used to refer only to data analysis, the scope is wider and the contributing elements of the field should be properly recognized. The purpose of this section is to discuss these four core areas in some detail, focusing on their specific issues and their contributions to data science.

The core is in close interaction with application domains that have the dual function of informing the appropriate technologies, tools, algorithms and methodologies that should be useful to develop, and leverage these capabilities to solve their problems. Some example applications are discussed in Section \ref{sec:apps}.

Data science application deployments are highly sensitive to the existing social and policy context and these influence both the core technologies and the application deployments. For example, what you can do with data differs in different jurisdictions. Data science applications and their deployments also have societal impact (both good and bad) and it is important to consider them within the data science ecosystem. One hopes that  the data science techniques, algorithms, considerations and deployments will impact development of evidence-based policies. Some of the social and policy issues related to data science are discussed in Section \ref{sec:soc}.

\subsection{Data Engineering}
\label{sec:de}

Data is at the core of data science; the type of data that is used is what is commonly referred to as ``big data''.  Data engineering component of data science has two main tasks: the \textit{management of big data} including the computing platforms for its processing, and the \textit{preparation of data} for analysis.

\subsubsection{Big Data Management}

There is no universal definition of big data; a working definition that is found in an NSF solicitation says the term ``refers to large, diverse, complex, longitudinal, and/or distributed datasets generated from instruments, sensors, Internet transactions, email, video, click streams, and/or all other digital sources available today and in the future.'' (quoted in ~\citep{Lane:2014tz}) This is likely as reasonable a definition as is possible. However, the more common characterization of big data is by its four identifying properties (what is called the ``four V's''): volume, variety, velocity, and veracity. 
 \begin{itemize}
  \item \textbf{Volume.} The datasets that are used in data science are commonly very large, typically in the petabyte range and with the growth of Internet-of-Things applications soon to reach zettabytes -- IDC's recent prediction is that the global data volume will reach 175 Zettabytes by 2025. To put this in perspective, Google has reported that in 2016, user uploads to YouTube required 1PB of \emph{new} storage capacity \emph{per day}. This was expected to grow exponentially, with $10\times$ increase every five years (so by the time of this writing, their daily storage addition may be 10PB). Facebook stores about 250 billion images  (as of 2018) requiring exabytes of storage. Alibaba has reported that during a heavy period on their 11/11 (similar to Black Friday in North America) in 2017,  320 PB of log data was generated in a six hour period as a result of customer purchase activity. Of course, volume is a moving target; over time what is considered large changes -- sometimes dramatically.
  \item \textbf{Variety.} Traditional (usually meaning relational) database management systems (DBMSs) are designed to work on well-structured data -- that is what the data schema describes. In big data applications, this is no longer the case, and multi-modal data has to be managed and processed. The data may include structured data, images, text, audio, video, microblog data such as tweets and others. It has been claimed that 90\% of  data generated today is unstructured. The big data systems are expected to be able to manage and process all of these data types seamlessly. Furthermore, in data science, the relationships among the variety of datasets are significant and important. Although individual datasets may not be large or comples, the relationships both within a single dataset and across multiple datasets may be complex.
  \item \textbf{Velocity.} An important aspect of big data applications is that they sometimes deal with data that is arriving at the system at high speed requiring systems to be able to process the data as they arrive. Facebook has to process 900 million photos that users upload per day; Alibaba has reported that during a peak period, they had to process 470 million event logs per second. These numbers do not normally allow systems to store the data before processing, requiring real-time capabilities.
  \item \textbf{Veracity.\index{veracity}} The data used by big data applications come from many sources, each of which may not be entirely reliable or trustworthy -- there could be noise, bias, inconsistencies among the different copies and deliberate misinformation.  This is commonly referred to as ``dirty data'' and it is unavoidable as the data sources grow along with the volume. It is claimed that dirty data cost upwards of \$3 billion annually in U.S. economy alone. Big data systems need to worry about \emph{data quality} and maintain their provenance in order to reason about their trustworthiness. Another important dimension of veracity is ``truthfulness'' of the data to ensure that the data is not altered by noise, biases or intentional manipulation. The fundamental point is that the data needs to be trustable.
 \end{itemize}

Managing of this data is challenging but critical in data science application development and deployment. These data characteristics are quite different than those that traditional data management systems are designed for, requiring new systems, methodologies and approaches. What is needed is a data management platform that provides appropriate functionality and interfaces for conducting data analysis, executing declarative queries, and sophisticated search. 

\begin{figure}%
    \centering
    \subfloat[\centering Using a data warehouse]{\label{fig:dw}{\includegraphics[width=0.47\linewidth]{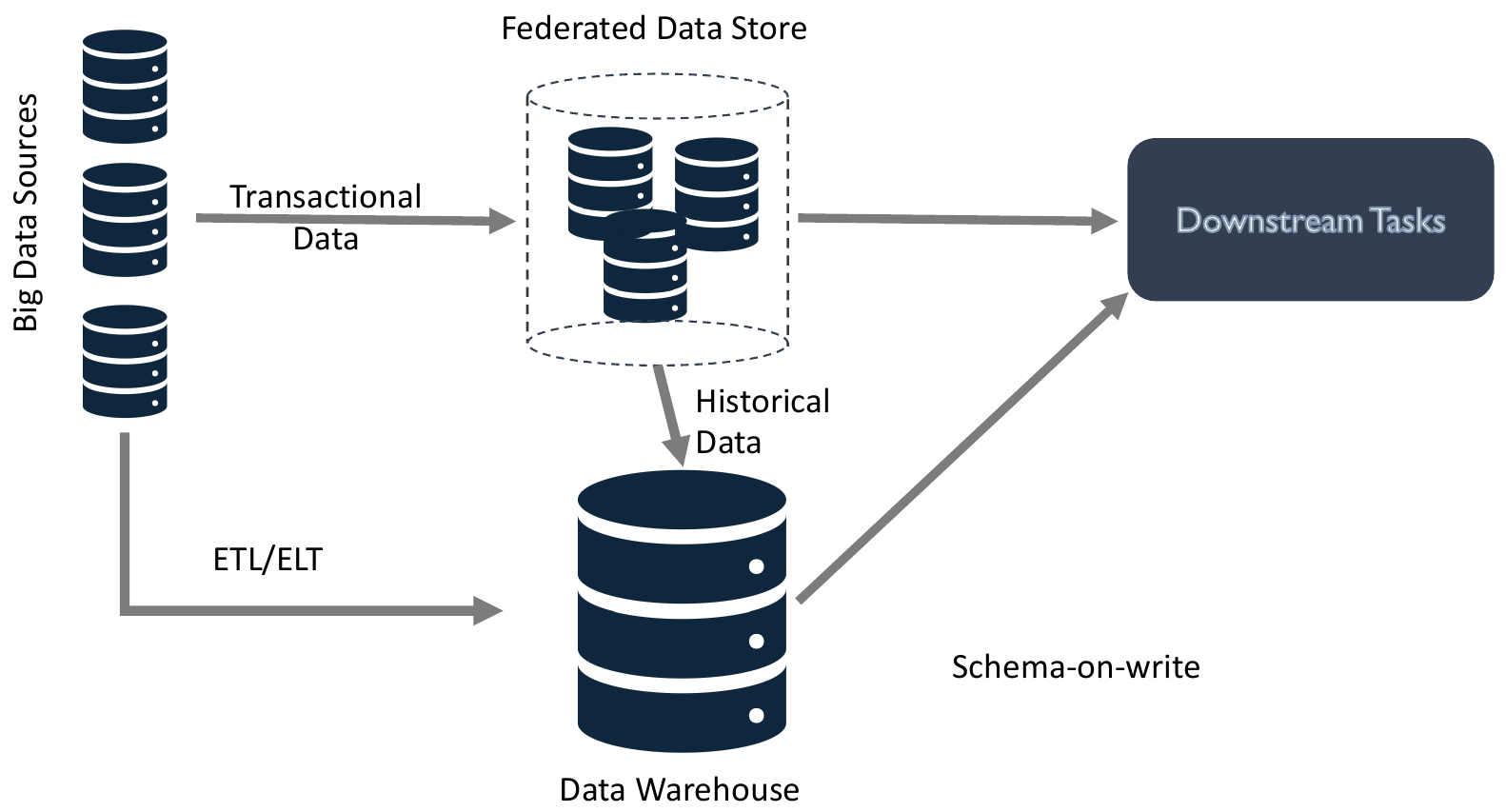} }}%
    \hfill
    \subfloat[\centering Using a data lake]{\label{fig:dl}{\includegraphics[width=0.47\linewidth]{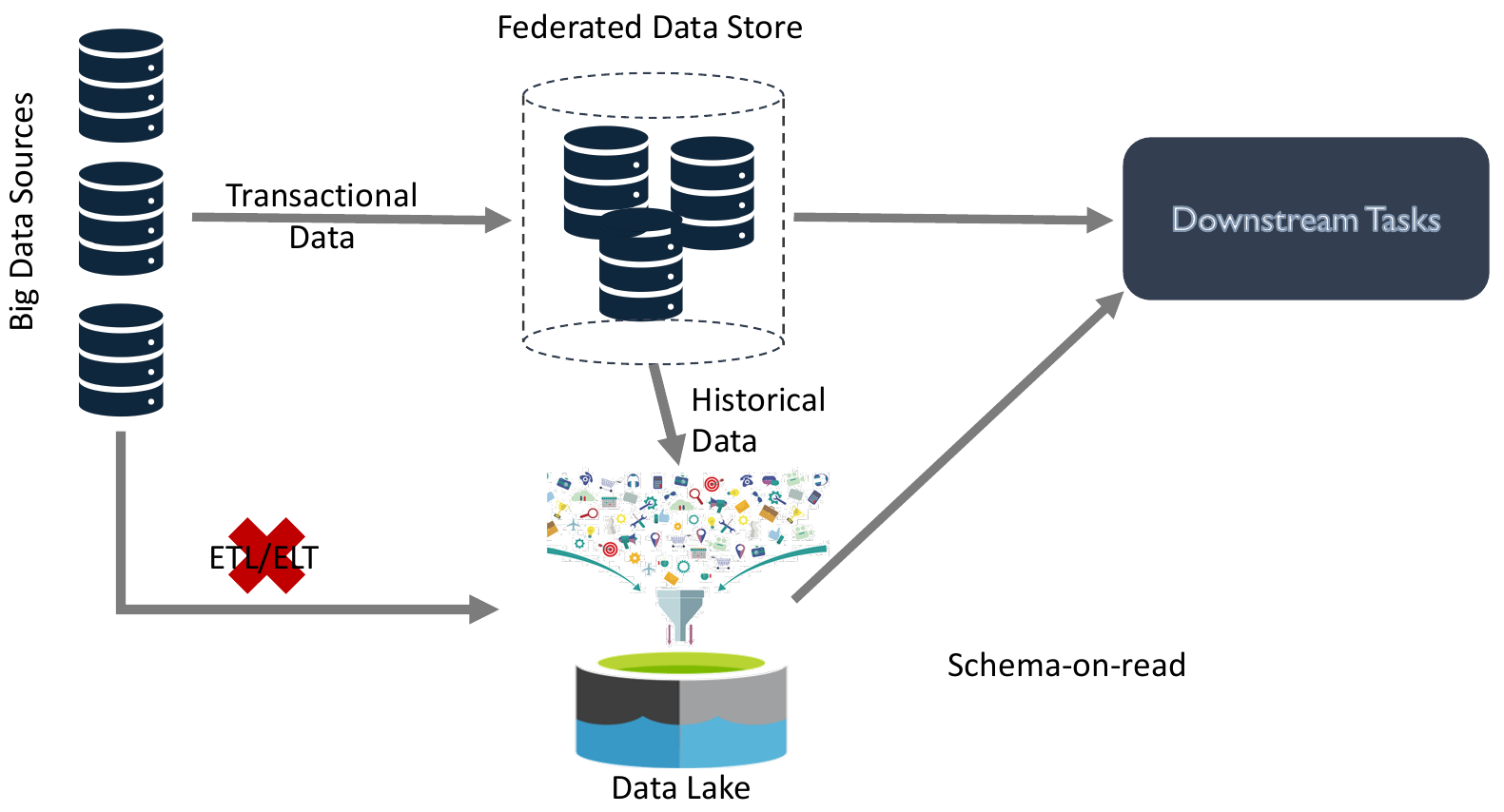} }}%
    \caption{Alternative integration architectures}%
    \label{fig:int-arch}%
\end{figure}

A data science platform needs to deal with multi-modal data that comes from different sources. Standardization efforts attempt to simplify data exchange and integration, but given the richness of the data, offering unified access to diverse data stores remains a challenge for a holistic and cross-disciplinary approach. There are two traditional ways of dealing with this issue, and in data science environments, these typically need to be combined. One is \emph{virtual integration} where data is maintained on the original sources, but mechanisms are provided to seamlessly access the data by providing an integrated view -- this is called the \emph{federated data store}. The second is \emph{physical integration} where data in loaded into a single store with a common format (i.e., schema) and accessed in the usual manner. The typical and long-studied approach to physical integration is building a \emph{data warehouse}. The source data is integrated into the data warehouse schema using extract-transform-load (ETL) or extract-load-transform (ELT) pipelines. This is commonly referred to as \emph{schema-on-write} as the data warehouse schema is enforced when data is written (loaded) into the data warehouse. This is demonstrated in Figure \ref{fig:dw}. 

A recent trend is to replace data warehouse in this processing architecture by a \emph{data lake} that incorporates data from many different sources and in many formats.  A data lake differs from a data warehouse in that the former  stores source data in its native format, using a big data store that can accommodate different formats such as polystores.  A data lake can be characterized as ``massive collection of datasets that (1) may be hosted in different storage systems; (2) may vary in their formats; (3) may not be accompanied by any useful metadata or may use different formats to describe their metadata; and (4) may change autonomously over time''~\citep{vldb19_NargesianZMPA19}. When the data is accessed by a query system or a data analysis tool, it is transformed into the format expected by the accessor; thus, they are \emph{schema-on-read} systems (Figure \ref{fig:dl}). It is easier (i.e., has a lower overheard) to create a data lake, but accessing data requires more work, because it requires transformations to multiple system interfaces. Data lakes typically support multi-workload data processing such as batch or realtime analysis as well as supporting ad hoc declarative query access. Building a physical integration system, be it a data warehouse or a data lake, is one of the technical challenges in data engineering.

Building and maintaining the computing platform (both hardware and software) that supports entire data science lifecycle as discussed in Section \ref{sec:lifecycle} is another challenge. The platform must be able to support a variety of models, enable transformations between them, allow for data integration, and offer a query language that all data practitioners, scientists or others, can use. This exceeds the current state-of-the-art, where individual systems are specialized toward a specific data model and require meta-models that allow for more fluent integration and seamless interoperability. In particular, the platform needs to provide multiple types of access: (1) simple transactional interfaces to update data, (2) powerful, declarative queries for online exploration, and (3) large-scale data mining and machine learning workloads for transformations and analysis. The focus of traditional database systems is to provide declarative access to the data through a query language, while big data analytics platforms, such as MapReduce~\citep{dean2010a,li:2013uq} and Spark~\citep{zaharia:2010,zahariabook:2016}, often follow a batch access. Given the increasing use of streaming data, these need to be adjusted for real-time processing using features of stream processing. As noted above, these requirements exceed what is possible with current data management/processing platforms. We discuss how the problem is addressed by engineering systems using different components in Section \ref{sec:arch}, but it is not clear whether this type of platform will continue to serve data science deployments as they become more sophisticated. This requires data to be transferred from database systems to the data analysis platform. 

\subsubsection{Data Preparation}
In addition to the big data storage, management and access concerns highlighted above, a fundamental aspect of data engineering is data preparation.  Applying appropriate analysis to the integrated data will provide new insights that can improve organizational effectiveness and efficiency and result in evidence-informed policies and actions. However, for this analysis to yield meaningful results, the input data must be appropriately prepared and trustworthy. The quality of the analysis model makes little difference; if the input data is not clean and trustworthy, then the results will not be of much value. The adage of ``garbage in, garbage out'' is real in data science.

Data preparation is typically understood to be the process of data acquisition, dataset selection, data integration, and data quality enforcement. We discussed data integration above, and data acquisition and dataset selection are discussed later in the paper when data science lifecycle is introduced. We focus on data quality in the remainder. 

Data quality is an essential element in making the data trustworthy for analysis. It addresses the veracity characteristic of big data. Data quality is considered mission critical in data science success, and constitutes a major portion of the data preparation effort for most organizations. Data quality has been identified as a major problem in industry: 89\% of executives believe that data quality issues impact the quality of customer service they provide~\citep{Experian2019}, only 33\% of senior executives have a high level of trust in the accuracy of their big data analytics~\citep{KPMG2016}, 59\% of executives do not believe their company has capabilities to generate business insights from their data\citep{Berez2016}. A Gartner study in June 2018 reports that poor data quality costs organizations an average of \$15 million a year~\citep{Gartner2018}; one expects the amount to have increased since then due to increasing deployment of data science applications. These quantify only the impact in industry; given data science applications in  medicine, health, environment (broadly defined), the criticalness of data quality should be obvious. One does not need to look further than the recent difficulties with Covid-related data to highlight the issue~\citep{Chakravorti:2022aa}.

Various attempts have been made to define and frame data quality issues. Data Management Association (DAMA) of UK has produced a report ~\citep{Dama:2013tp} that identifies six fundamental dimensions of data quality. These were later expanded to 60 dimensions twelve of which were identified as the most important ~\citep{Black:2020tl}. There are even more extensive ones, with one study identifying 179 dimensions and summarizing them in 15 ~\citep{Wang:1996tt}. These studies are summarized well by Sattler \cite{Sattler2009}, but it is useful to look briefly at the shorter list of six to understand the scope of concerns.

\begin{itemize}
    \item \textbf{Accuracy} - How well does a piece of information reflect reality?
    \item \textbf{Completeness} - Does the data fulfill your expectations of what’s comprehensive?
    \item \textbf{Consistency} - Does data stored in one place match relevant data stored elsewhere?
    \item \textbf{Timeliness} - Is your data available when you need it?
    \item \textbf{Validity} - Is data in a specific format, does it follow business rules, or is it in an unusable format?
    \item \textbf{Uniqueness} - Is this the only instance in which this data appears in the database?
\end{itemize}

The definition of these (and other) dimensions need to be complemented by a definition and implementation of metrics to measure them. Despite significant commercial and academic research and tool development  for data quality management, it does not seem that metrics have been widely implemented ~\citep{Ehrlinger:2019wp}.

An important vehicle of data quality and data trustworthiness is metadata and metadata management. Dictionary definition of metadata is ``data about the data.'' The topic is broad, but one particularly important metadata that deserves mention is \emph{provenance}, which tracks the source of the original data, changes to the data, and changes to its properties (e.g., storage location). Provenance information is essential for ensuring data validity and reproducibility of analyses. One challenge is to quantify, and then implement, the expressiveness and granularity of the provenance model. While a highly expressive provenance language describing how data is extracted conveys useful information for downstream analysis, maintaining and reasoning about these provenance statements can be expensive and may not be available in most situations. Developing and instituting the appropriate system and tool support for managing provenance, and tracking data as it goes through the processing pipeline (i.e., data lineage) is another challenge. Propagating provenance information across heterogeneous systems is also a major system implementation challenge that is crucial for end-to-end data governance and curation. 

A very important aspect of data quality is data cleaning ~\citep{Ilyas:2019aa}, which cuts across all of the six quality dimensions discussed above. When data from multiple sources are integrated (logically or physically), there are bound to be inconsistencies, errors in data as well as missing information. Data cleaning is the process of detecting and correcting (or removing) corrupt or inaccurate records from a dataset. The development of techniques for this task rely on both machine intelligence and on the specialized knowledge of domain experts. Classical inference and machine learning methods, such as naïve Bayes and decision trees, can work well, but the challenge is in choosing the most important attributes for implementing these methods. Deep learning shows some promise in avoiding heavy feature engineering, whereas active learning techniques aim to identify the most effective training data with input from expert users. This line of research will build data cleaning platforms that can be easily adapted to individual data user’s needs and can provide non-destructive and knowledge-based data cleaning. Using statistical models for the identification and correction of outliers, anomalies, and inconsistencies can strengthen the performance of subsequent algorithms for analysis.

\subsection{Data Analytics}
\label{sec:da}

The second component of the data science ecosystem is data analytics. Data analytics is the application of statistical and machine learning techniques to draw insights from data under study and to make predictions about the behaviour of the system under study. 
%This is a generalization of the definition  provided by Tukey in early days that focused more on the statistical analysis while restricting its scope by removing activities we now identify as part of data engineering: ``Procedures for analyzing data, techniques for interpreting the results of such procedures, ways of planning the gathering of data to make its analysis easier, more precise or more accurate, and all the machinery and results of (mathematical) statistics which apply to analyzing data.'' \cite{Tukey:1962ue}

There is usually a discussion of the respective roles of statistics and machine learning in data analytics that will be discussed in Section \ref{sec:owner}. It is safe to say that the lines of demarcation between these two disciplines have never been clear and is increasingly blurred. Perhaps a reasonable separation is that ``statistics draws population inferences from a sample, and machine learning finds generalizable
predictive patterns." \citep{Bzdok:2018wt}. The authors state ``Classical statistical modeling was designed for data with a few dozen input variables and sample sizes that would be considered small to moderate today. In this scenario, the model fills in the unobserved aspects of the system.
However, as the numbers of input variables and possible associations among them increase, the model that captures these relationships becomes more complex. Consequently, statistical inferences
become less precise and the boundary between statistical and ML approaches becomes hazier.'' Without getting unnecessarily hung up with the respective roles of each discipline, this may be as clear an identification of their complementary roles as is possible.

A first-level distinction in data analysis is made between \textit{inference} and \textit{prediction}. Inference is based on building a model that describes a system behaviour by representing the input variables and relationships among them. Prediction goes further and identifies the courses of action that might yield the ``best'' outcomes. This classification can be made more fine-grained by identifying four different classes:

\begin{itemize}
    \item \textbf{Descriptive.}  The question that descriptive analysis tries to answer is ``what happened?'' or ``what does the data tell us?'' It is retrospective in that it looks at historical data in an attempt to answer the ``what'' question. One particular version of descriptive analytics is \textit{exploratory analysis} \citep{Tukey:1977uy}. In data science, most data analytics is targeted -- there is a problem definition and analytics is targeted to that definition (more on this in Section \ref{sec:lifecycle}). This is true even in answering the ``what'' question, because there is usually a context for the question and the task is to build a model to explain what happened. This has been called \textit{confirmatory} [\textit{descriptive}] \textit{analytics} \citep{Tukey:1980wx}. In some cases formulating this question properly may not be possible or easy; in those cases exploratory analysis where the intent is discovering what the data reveals is useful. Modern exploratory analysis typically depends on heavy data visualization and is sometimes called \textit{visual analytics}.
    
    \item \textbf{Diagnostic.} This is also a retrospective analysis, but it goes beyond descriptive by asking the question ``why has that happened?'' It looks at what the data suggests are the reasons for the observed phenomenon. This requires drilling down into the data to determine the correlation and causation relationships between variables in the model. Correlation implies that the variables in question change together, while causation requires that the movement of a set of variables determine the movement of the other variable(s).     It is usually easier to determine correlation between variables than causation, and the phrase ``correlation is not causation'' is now well-ingrained in students of data analytics. Although causation is better, correlation can still reveal interesting diagnostic information. 
    
    \item \textbf{Predictive.} Predictive analysis differs from descriptive and diagnostic in that it is a forward-looking analysis of historical data. It provides calculated predictions of what is likely to happen. Predictive typically uses machine learning, data mining, and predictive statistical modeling. Predictive analysis is important in data science applications such as fraud detection where past transactions are analyzed to predict whether the current transaction is fraudulent.
    
    \item \textbf{Prescriptive.} Prescriptive analysis is also forward-looking, but goes further by recommending courses of action. The question that is answered is ``what should be done?" It takes the information that has been predicted and prescribes calculated next steps to take. Predictive analysis typically requires heavy machine learning usage, simulation and complex event analysis. A typical data science application that uses prescriptive analytics is recommendation systems where users are provided with recommendations as to which items to buy, which movies to watch, and, in general, which actions to take. 

\end{itemize}

\begin{figure}
    \centering
    \includegraphics[width=0.5\linewidth]{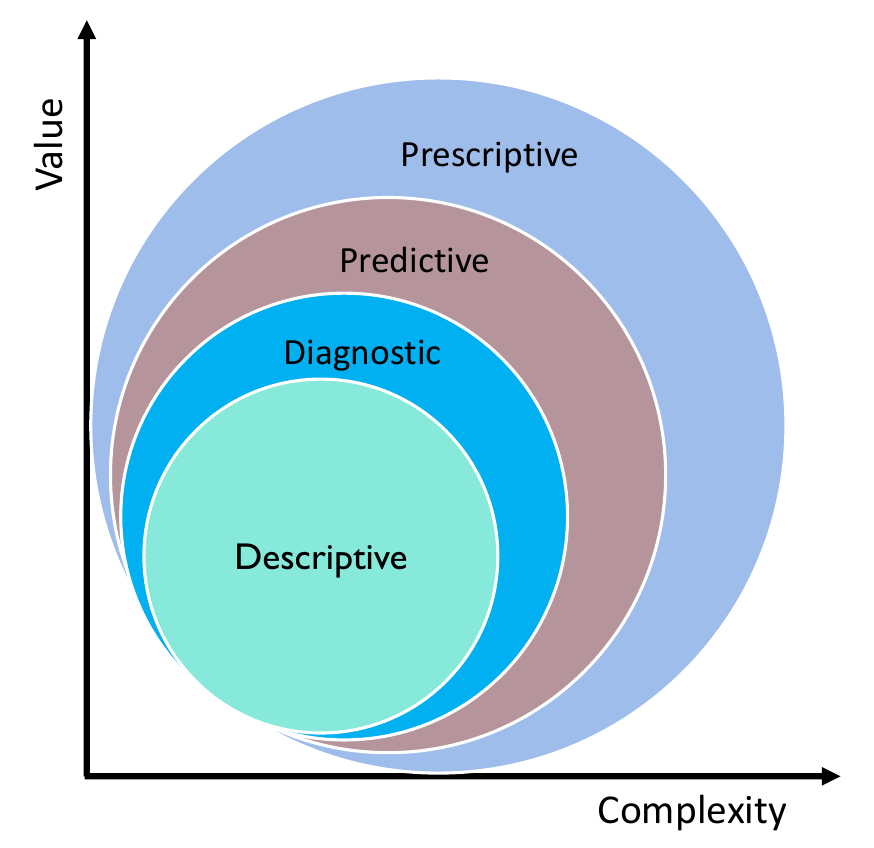}
    \caption{Four Types of Data Analytics and their Relationships. From  \cite{Maydon07}}
    \label{fig:analytics}
\end{figure}

Predictive and prescriptive analytics together are usually called advanced analytics, because of techniques that are employed that go beyond  statistical modeling. The relationship among these analytics types are usually evaluated along two dimensions: complexity and value (Figure \ref{fig:analytics})~\citep{Maydon07}. Going from descriptive to prescriptive, analysis becomes far more complex, but the value derived from it also substantially increases.

There are six data analysis tasks (methods) that are commonly used in data science \citep{Fayyad:1996vz,Kelleher:2018tv}:

\begin{itemize}
    \item \textbf{Clustering.} Clustering finds meaningful groups or collections of data based on the ``similarity'' of data points -- data points in the same cluster are more similar to each other than they are to data points in other clusters. It is an unsupervised learning technique where data items are not labeled and the number of groups (clusters) is not fixed apriori. Some example applications are market segmentation, social network analysis, grouping of search result, and medical imaging.
    
    \item \textbf{Outlier detection.} Also called \textit{anomaly detection}, it refers to identification of rare data items in a dataset that differ significantly from the majority of the data. Typical applications are fraud detection, medical problems (e.g., finding cancerous tissues), and finding errors in text.
    
    \item \textbf{Association rule learning.} This method discovers interesting relationships between variables in a large dataset. Each rule indicates that if an antecedent is true, then the consequent is also true. The typical application example is market basket analysis which discovers rules linking a customer's likelihood of purchasing a particular item (consequent), if she has purchased  another item (antecedent). The \textit{support} and \textit{confidence} of a rule is computed to determine if it is statistically supported by the available data.
    
    \item \textbf{Classification.} Classification finds a function (model) that places a given data item in one of a set of predefined classes. It is a form of supervised learning where  the  structure learned over a dataset is generalized to a new dataset. Some examples where classification is used are in financial applications (e.g., determining whether someone is loan-worthy or not), object identification in images, whether or not an email is junk/phishing attack.
    
    \item \textbf{Regression.} Regression finds the function that relates one or more independent variables to a dependent variable. The function determines how the dependent variable changes when the independent variables change. This is similar to correlation discussed above with the difference being that ``correlation measures the \textit{strength} of an association between [the independent and dependent] variables, regression quantifies the \textit{nature} of the relationship.'' \citep{Bruce:2020ts} Regression is also a supervised learning technique. Some example applications may be predicting sales based on advertisement expenditures, and public health applications that relate spread of illness to certain measure(s).
    
    \item \textbf{Summarization.} Summarization creates  a more compact representation of the dataset. This could be either in the form of a tabular report or a visualization. For example, a histogram or table of sales over the months of a year is a summarization of the sales data.

\end{itemize}

Relating these tasks (methods) with the classes of analysis discussed earlier, clustering, association rule mining, and summarization are techniques that are typically used for descriptive analysis, while classification, regression and outlier detection are more suitable for predictive analysis. 

\begin{figure}%
    \centering
    \subfloat[\centering Batch analytics]{\label{fig:anal-batch}{\includegraphics[width=0.3\linewidth]{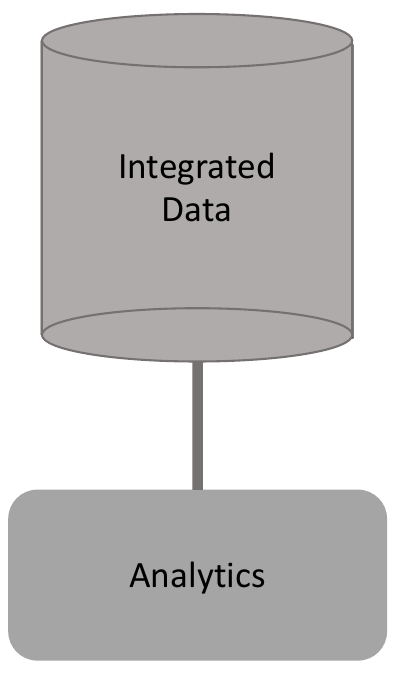} }}%
    \hfill
    \subfloat[\centering Federated analytics]{\label{fig:anal-federal}{\includegraphics[width=0.3\linewidth]{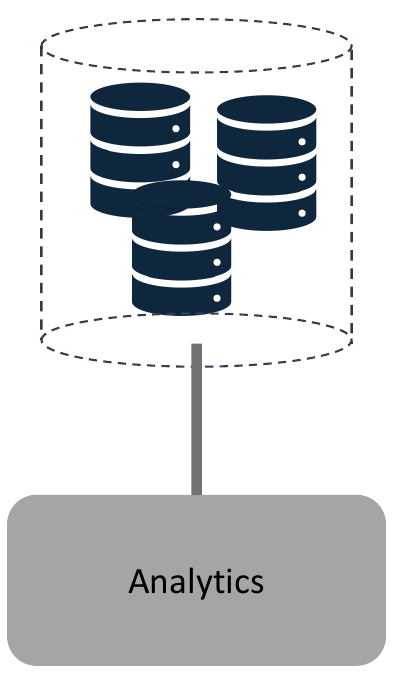} }}%
    \hfill
    \subfloat[\centering Realtime analytics]{\label{fig:anal-real}{\includegraphics[width=0.3\linewidth]{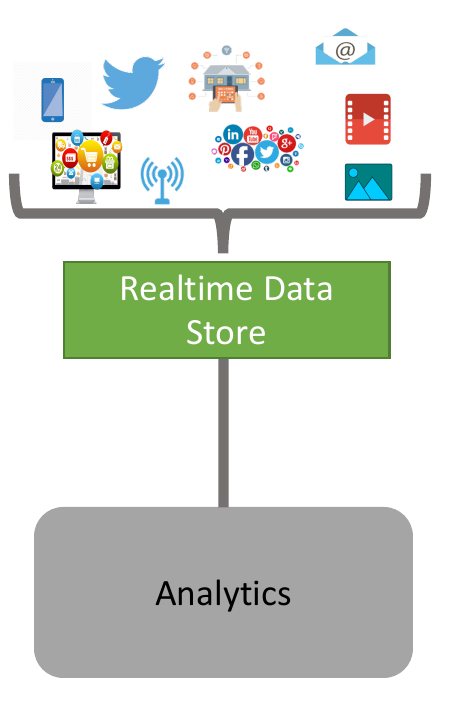} }}%
   \caption{Alternative analytics approaches}%
    \label{fig:alt-analytics}%
\end{figure}

There are alternatives ways to architect data analysis, following the data integration discussion in the previous section. One possibility, and probably the most common and well-understood one is to physically integrate the data on which \emph{batch analytics} is performed (Figure \ref{fig:anal-batch}). Many statistical and machine learning techniques have targeted analysis of a single dataset in batch mode. However, it gives rise to a number of difficulties. First, scaling to very large integrated datasets may be difficult, resulting in re-partitioning the data for scale-out processing. The issues of scale-up versus scale-out processing are topics of ongoing debate\footnote{An example of this debate for graph data is provided in  ~\citep{Lin:2018aa,Salihoglu:2018aa}.}. If scale-out is followed, this results in \emph{distributed analytics}. Second, integrating data in this fashion increases the data protection challenges as discussed in the next section. Finally, and perhaps most importantly, the owners of the data may not be willing to give up control for physical integration -- especially if subsequent repartitioning may require data to be located on servers with which they do not have an established trust relationship. They may be willing to share access to data, but not the actual data. 

These issues have given rise in recent years to \emph{federated analytics}~\citep{Kairouz:2021ux,Yang:2019tj} where analysis is performed on logically integrated datasets (Figure \ref{fig:anal-federal}). This has given rise to discussions of entire federated data science scenarios~\citep{Mansour:2022ud}. One advantage of these systems is that, with careful design, it is possible to increase availability of data for analysis as the failure of one or a few of the data stores would not stop the entire analysis activity. 

As discussed earlier, an important data source in data science is streaming data. In this case, \emph{realtime analytics} must be considered as data flows continuously (Figure \ref{fig:anal-real}). Realtime analytics is particularly difficult given that most analysis algorithms are computationally heavy and usually require multiple passes over the dataset, which is not normally possible over streaming data. System techniques such as windowing (within the context of a transient realtime data store) are usually exploited to perform analytics over small batches with provisions taken to catch patterns that may span multiple windows. There have been approaches to create streaming data warehouses~\citep{golab:2010fj} where data is captured in the backend, and realtime processing/analytics is complemented by batch analytics on the captured data. An issue that arises prominently in streaming data analysis is \emph{concept drift}, which refers to changes that occur over time to the underlying data characteristics that may require adjustments to the models.

In a data science project, the issue of selecting the appropriate technique(s) for the task and how they can be leveraged is an important consideration. Given the societal impact of data science applications and deployments, the explainability of the analysis results is equally important.

\subsection{Data Protection}
\label{sec:protect}

%\improvement[inline]{\normalsize This is double the size of other sections in ecosystem.\newline Either shorten this or expand others.}

Data science's reliance on large volumes of varied data from many sources raises important data protection concerns. The scale, diversity, and the interconnectedness of data (for example, in online social networks) requires revisiting the data protection techniques that have been mostly developed for corporate data~\citep{Bertino:2018wo,Moura:2019ji}. 

It is customary to discuss the relevant issues under the complementary topics \emph{data security} and \emph{data privacy}. The former protects information from any unauthorized access or malicious attacks, while the latter focuses on the rights of users and groups over data about themselves. Data security typically deals with data confidentiality, data integrity, access control, infrastructure security, and system monitoring, and uses technologies such as encryption, trusted execution environments, and monitoring tools. Data privacy, on the other hand, deals with privacy policies and regulations, data retention and deletion policies, data subject access requirement (DSAR) policies, management of data use by third parties, and user consent. Data privacy normally involves privacy enhancing technologies (PETs). Although research on these topics are usually isolated, it is helpful to take a holistic and broader view, hence the term \emph{data protection} is likely more appropriate and informative.

The characteristics of data used in data science pose unique challenges. Data volumes make the enforcement of access control mechanisms more difficult and the detection of malicious data and use more challenging. The numerosity and variety of data sources make it possible to inject mis-/dis-information, skewing the analysis results. Data science platforms, as discussed in Section \ref{sec:arch} are, by necessity, scale-out systems that increase the possibility of infrastructure attacks. These environments also increase the potential for surveillance. The variability and potentially high numbers of end users, and, in many data science deployments, the need for openness for sharing  analysis results and for bolstering the analysis, opens  the possibility of data breaches and misuse. These factors seriously increase the threats and the attack surface. Therefore, protection is required for the entirety of data science lifecycle (Section \ref{sec:lifecycle}) from data acquisition to the dissemination of results, as well as for secure archiving or deletion. An implicit goal of data science is to gain access to as much data as possible, which directly conflicts with the fundamental least-privilege security principle of providing  access to as few resources as possible. Closing this gap requires careful redesign and advancement of security technologies to preserve the integrity of scientific results, data privacy, and to comply with regulations and agreements governing data access. Techniques that have been developed for privacy-preserving data mining are examples of this consideration.

\begin{figure}
    \centering
    \includegraphics[width=0.8\linewidth]{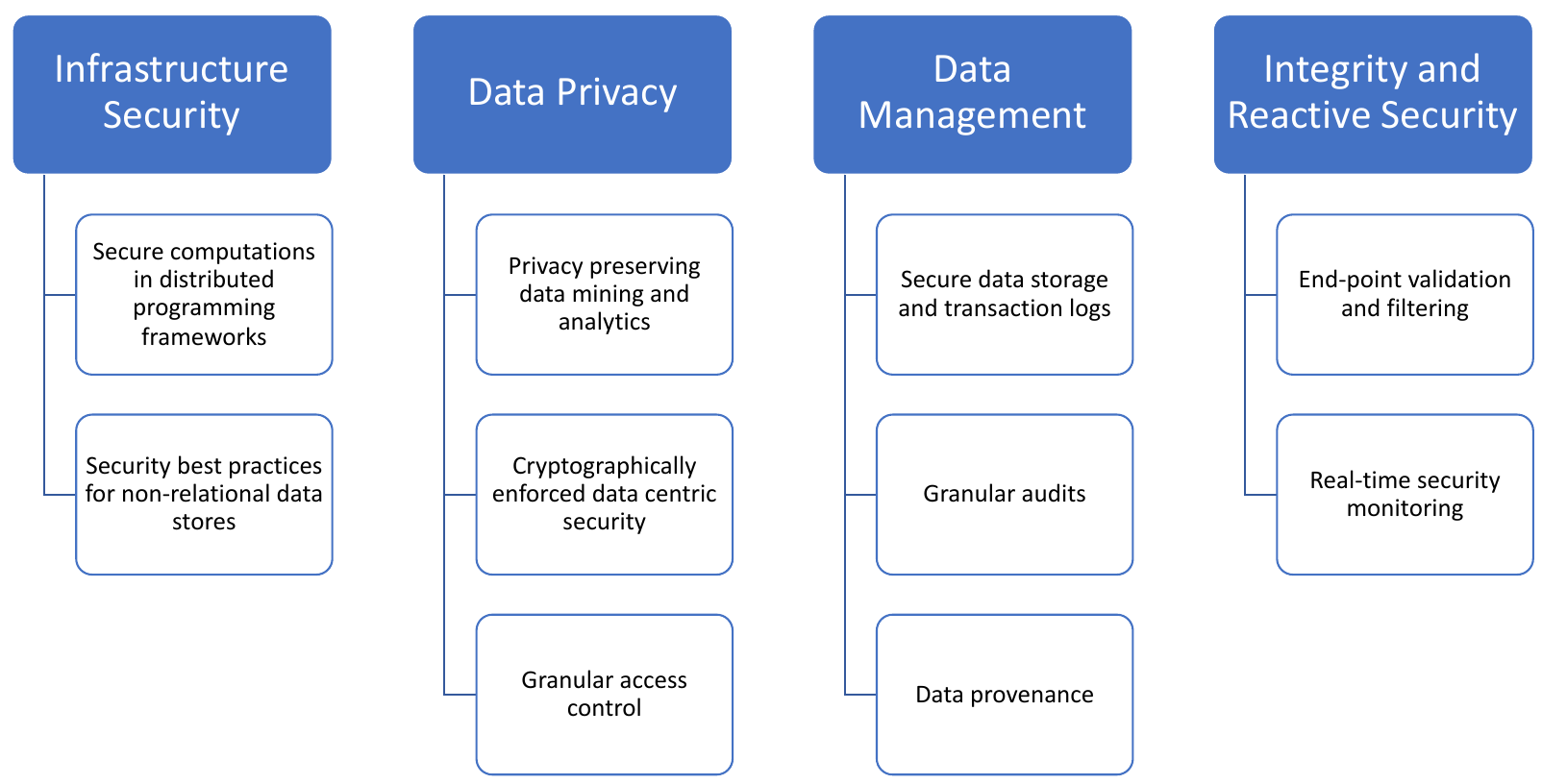}
    \caption{Cloud Security Alliance Identified Challenges. From  \citep{CSA:2013aa}}
    \label{fig:csa}
\end{figure}

This view argues for a move from the notion of protecting data to a broader cybersecurity viewpoint where protection is considered at the platform, network, software levels in addition to data. Cloud Secure Alliance has identified ten major challenges (Figure \ref{fig:csa}) and categorized them in terms of the protection that should be provided~\citep{CSA:2013aa}.

%\begin{itemize}
%\item Infrastructure security
%\begin{itemize}
%\item Secure computations in distributed programming frameworks
%\item Security best practices for non-relational data stores
%\end{itemize}
%
%\item Data privacy
%\begin{itemize}
%\item Privacy preserving data mining and analytics
%\item Cryptographically enforced data centric security
%\item Granular access control
%\end{itemize}
%
%\item Data management
%\begin{itemize}
%\item Secure data storage and transaction logs
%\item Granular audits
%\item Data provenance
%\end{itemize}
%
%\item Integrity and reactive security
%\begin{itemize}
%\item End-point validation and filtering
%\item Real-time security monitoring
%\end{itemize}
%
%\end{itemize}

\subsubsection{Security Concerns}

As noted above, security deals with the protection of data. A helpful characterization is given in the U.S. Federal Information Management Act of 2002 that establishes the concerns as confidentiality, integrity and availability. \emph{Confidentiality} refers to restricting access and disclosure, \emph{integrity} refers to ensuring that the data is correct and protected against improper modification and destruction, and \textit{availability} refers to timely and reliable access.  Integrity has been generalized to data trustworthiness that encompasses data correctness and the reliability of sources, raising the issues of data quality and data provenance discussed in Section \ref{sec:de}. 

All of these concerns are codified in a \emph{security policy} that specifies the allowable and non-allowable actions and security expectations. These policies are then implemented through a set of \emph{security mechanisms}. In data science deployments, data comes from multiple sources that may be physically or logically integrated. Each data source might have different policies that are established autonomously, and it becomes necessary to reconcile these and to negotiate which ones might be applied~\citep{Bertino:2018wo}. This is also called \textit{multilateral security}~\citep{Pfitzmann:2006ue}.

There are different ways to frame the discussion of security issues. One is the combination of the three dimensions of information security as discussed above (confidentiality, integrity, availability) and the critical phases of data science lifecycle. These are orthogonal dimensions and at each data point in the two-dimensional space, there are security issues. Another alternative is to consider security of data-at-rest (i.e., when it is stored), data-in-motion (i.e., when it is being transmitted through the system), and data-in-use (i.e., while it is being processed); finally, it is possible to consider the issues at various layers of the system: infrastructure, software and data security~\citep{Tellenbach2019}. In the following, the issues are highlighted following the last alternative. 

\paragraph{Infrastructure security.}

Data science deployments run on complex and varied computing and networking environments; the specific architecture will be discussed in Section \ref{sec:arch}, but it is sufficient at this point to consider that the infrastructure includes large and multiple data storage systems, computing systems on which application code executes, and the networking that connects all of these components. Sometimes, the setup might be in a single data center, but more often they involve multiple data centers that might be geographically distributed. This environment has a very large attack surface; the perimeter of the computing infrastructure needs to be defined and protected.

There are various aspects of infrastructure security  and they relate to confidentiality, integrity and availability aspects of information security. An important aspect is preventing unauthorized  access to the physical infrastructure. The security mechanism used for this purpose is \emph{physical access control} (PAC)  (different than data access control discussed below) that involves deployment of tools such as keys, electronic access cards, etc. 

An important aspect of infrastructure security is network and communication security. The network (communication infrastructure) can be subject to the following types of threats: (a) interception of the transmitted message or data by an unauthorized person; (b) interruption of the communication because the network has become unavailable, unusable or destroyed; (c) unauthorized modification of the transmitted message or data; and (d) fabrication of data or message that does not exist~\citep{Maarten:2017aa}. These threats, if successful, harm the confidentiality, integrity and availability of the communication and data. A malicious third party might attack the network by (a) eavesdropping to listen to network traffic without authority, (b) masquerading as an authorized party by assuming an identity, (c) tampering with messages by intercepting them and changing their contents before delivering it to the intended recipient (man-in-the-middle attack), (d) capturing and storing messages and replaying them at a later time, and (e) flooding the network and the computing equipment connected to it by extreme number of messages to overwhelm the system (denial-of-service attack). It is easy to see how these would harm confidentiality of messages and data, the integrity of the data, and system availability.

One security mechanism that is required to secure networking infrastructure involves \emph{logical access control} (LAC). LAC  typically includes three complementary issues: \emph{authentication} to verify the claimed identity of a party (user, client, server, etc),    \emph{authorization} to check whether the requesting party is authorized to perform the requested action, and \emph{auditing tools} to trace, retroactively, who has taken what action and when. Authentication is typically implemented through challenge-response protocols. The two parties who wish to communicate share a key and the initiator sends a challenge to the other party that can only be resolved if the other party knows the common key among them.

The second security mechanism that needs to be employed for network security is encryption that is important for message (and data) confidentiality. This topic is discussed in detail below under data protection. For message integrity, \emph{cryptographic hashing} can be used. The message is hashed by a one-way function that generates an encrypted message that is then transmitted over the network. An unauthorized third party would not be able to read the message even if it is captured. Since it is a one-way function, it is not possible to recover the original message, and that differentiates cryptographic hashing from encryption.

The final security mechanism that is useful for message integrity is digital signatures. The concern here is that the message may be captured and modified. Digital signatures uniquely tie the message content to the sender. Public-key cryptographic methods discussed below are used to generate signatures.

The final aspect of infrastructure security involves availability if any component of the system fails or is unavailable. The basic approach to address this issue is \emph{replication} -- replication of computing components, replication of network connections, and replication of data. As part of data replication, data is usually stored at a different site to guard against disasters that may make data sources unavailable.

\paragraph{Operating system security.}

Similar to the hardware platforms for data science deployments, the software platforms are also complex and varied. This variability makes it difficult to define broadly applicable access control policies. Some of the existing platforms allow arbitrary user code to be executed that further complicate definition of fine-grained access control~\citep{Bertino:2018wo}.

These issues are handled at the operating system level. An important aspect of operating system security is protecting data-in-use. An emerging technology for this purpose is a \emph{trusted execution environment} (TEE), which is ``a secure, integrity-protected processing environment, consisting of memory and storage capabilities''~\citep{Sabt:2015ut}. It provides an isolated and secure execution kernel for secure code for data-in-use. TEE is a promising technology, but they may not prevent \emph{access pattern leakage} as a result of a compromised operating system~\citep{Zheng:2017wl,Xu:2015td}.

\paragraph{Software security.}

Software security deals with issues at the software levels above the operating system, such as middleware software, utility software, and application software. Within this context, the issue of controlling access to databases has been studied for some time -- a common issue is what is called ``SQL injection.'' The problem arises in applications where SQL queries to the database require user input. A malicious user can insert a text string as part of this input that results in retrieving data that the user is not authorized to see. Solutions to safeguard against SQL injection have been developed, but not all data science applications  employ relational database systems. The general class of data management systems called NoSQL have gained favour to manage the complex data in these applications, and each of these systems have their own interfaces, increasing the difficulty of providing general solutions.

%The analysis software that is used can also leak information that is not intended for release. 

\paragraph{Data security.}

Data security fundamentally addresses confidentiality, which ``deals with protecting against the disclosure of information by ensuring that the data is limited to those authorized or by representing the data in such a way that its semantics remain accessible only to those who possess some critical information.''~\citep{Paulsen:2019te}. The working assumption is that data resides on untrusted data servers while the access is either by applications that run on a trusted application server or by users on trusted clients. In this context, the questions that arise relate to the target of confidentiality: the data may need to be confidential and/or the application accesses/user queries may also need to be confidential~\citep{Moura:2019ji}. %In the following discussion, to simplify matters, the focus is only on data encryption but the concerns are similar.

An important technology for ensuring confidentiality is data encryption, which scrambles raw data (\textit{plaintext}) into encrypted data (\textit{cyphertext}) that cannot be understood by unauthorized third parties even if it is intercepted. Authorized parties hold a key that allows them to dycypher dat afor reading. What the keys are and how they are managed give rise to different classes of encryption algorithms and techniques.

Typical workflow for accessing encrypted dat is to decrypt it first. The data is requested from the provider, which sends it in encrypted format to the trusted requestor (an application or a user) where it is decrypted and processed. This can have high overhead in data movement and the preference is to push processing to the data stores, which requires the ability to process data in encrypted form.  Techniques such as \emph{homomorphic encryption}~\citep{Will:2015te,Sen:2013vz} and full homomorphic encryption~\citep{Gentry:2009ua} allow computation over encrypted data. However, the overhead of full homomorphic encryption is prohibitively high for real deployments, resulting in partial homomorphic encryption techniques that allow certain computations.

The data management community has long been interested in being able to run declarative queries (i.e., SQL) over encrypted data, e.g. ~\citep{Agrawal:2004up}. More recent attempts use a suite of practical encryption techniques and decompose queries to components that can be efficiently executed over encrypted data. The results are transmitted (in encrypted form) to the trusted requestor and further computation may be performed on the trusted requestor site following decryption of the data. Examples include CryptDB~\citep{Popa:2011wn} for transactional database workloads and Monomi~\citep{vldb13_tu_processing_2013} for analytical workloads.  

Another important consideration in providing confidentiality is \emph{access control} that concerns who can access which data.  It typically includes two complementary issues: \emph{authorization} of those who can access the data and \emph{authentication} of a requestor to determine access rights. Access control can be provided at different granularities; for example it can be on the entire dataset or on individual units (however they are defined) of the dataset.  There is an obvious connection between access control and privacy regulations that may dictate access rules to data. Although many access control methods have been defined for data management systems, their interaction with these regulations at a large scale are not well understood~\citep{Kantarcioglu:2019wb}. In data science deployments, since the data volume is very large, if fine-grained access control is required, administering authorizations manually is not feasible and automated tools need to be developed~\citep{Bertino:2018wo}.

\subsubsection{Data Privacy}

Data privacy concerns the rights of users and groups (data subjects) over data about themselves. In that sense, there is a natural relationship between confidentiality and privacy; privacy requires confidentiality, but it ``has additional issues deriving from the need of taking into account requirements from legal privacy regulations, as well as individual privacy preferences''~\citep{Bertino:2018wo}. Thus, the data privacy issues are extensions of confidentiality, and can be considered under three headings: (1) rights of subjects to control what data is captured and stored about them by those that create or store data (data providers); (2) the rights of subjects to control what the data providers can do with that data; (3) the rights of subjects to control what third parties can see and do with the data. Of particular concern is control over the capture, storage and release of personally identifiable information (PII), which is sensitive information. 

Many jurisdictions have enacted laws or policies that establish the legal framework governing data privacy and data science deployments operate within this framework. There are differences in the rights each jurisdiction awards the parties, and this is an issue in deployments in data science that span multiple jurisdictions.

Data holders and creators have an incentive to collect as much data as possible and be as unencumbered as possible in its use. Data subjects, on the other hand, typically have an interest in controlling the use of their own sensitive and identifying data. A fundamental challenge is to balance these conflicting goals, allowing subjects to have meaningful and sustained control over data about themselves while maintaining meaningful utility, including data analysis. The tradeoff between privacy and utility is a fundamental challenge. This tradeoff, along with the heterogeneity of ownership of the underlying computing infrastructure and the uncertainty of the ownership of data are major issues. They raise  questions about the possibility for subjects to maintain control over their data throughout the entire data science lifecycle (see Section \ref{sec:lifecycle})~\citep{Simo:2015to}. It is important to think of the privacy issues as they arise from data collection to storage to analysis and dissemination of results and data~\citep{Jain:2016us}.

Privacy concerns arise in data science because of the data scale, the diversity and numerousity of data sources, and how data is used. If data from multiple sources are physically integrated, this increases the possibility of revealing unintended information since interrelationships are made explicit through tight integration. Therefore, interest in federated analytics has increased and privacy-preserving techniques have been developed~\citep{Yin:2021vp}. However, even these techniques may not be sufficient to address privacy concerns. Data science is all about creating linkages among data that reveal new insights. These linkages, even in a federated setting, are potential sources for privacy leaks. 

Data analysis draws inferences among data, and data privacy concerns arise when sensitive information can be inferred from metadata or other data and not from the sensitive data itself. This has been called the ``inference problem''~\citep{Farkas:2002vn}, and has been studied within the database and data mining communities. Solutions, again, raise the tradeoff  between privacy and utility. Similarly, data science processing perform considerable aggregation of data, which might give a false sense of security that individual data is hidden in the aggregates. However, it has been well-established that aggregates, coupled with other data sources, can cause sensitive information to be leaked.

The set of technological solutions developed to protect sensitive information in data-at-rest are called \emph{privacy enhancing technologies} (PET). An important component of PETs are techniques that obfuscate the sensitive information in data; this has been referred to as \emph{content privacy}~\citep{Yu:2016ur}. The simplest and earliest technique is \emph{anonymization} of the sensitive information that can take one of three forms~\citep{Mortazavi2015}: (1) removal or omission of sensitive information; (2) pseudonymization that involves replacing sensitive information by some code; and (3) grouping, aggregation or classification of the sensitive information. Simple anonymization is not sufficient and anonymized data is still open to re-identification attacks. More elaborate anonymization methods have been proposed to counter re-identification, such as k-anonymity~\citep{sweeney2002}, $l-$diversity~\citep{machanavajjhala2006} and $t-$closeness~\citep{icde07_4221659}.

The current state-of-the-art in obfuscating sensitive information is \emph{differential privacy}~\citep{Dwork:2017wc,Dwork:2014vw}. Differential privacy is a statistical approach that quantifies information leakage and guarantees that is it below a threshold $\epsilon$. The intuition behind this approach is the following: if the results produced by an algorithm that is run on one dataset and on a modified dataset that removes information about one subject are not statistically distinguishable (measured by $\epsilon$), then the algorithm is said to satisfy $\epsilon-$differentially privacy~\citep{vldb16_Machanavajjhala:2016}. This is achieved by adding noise proportional to $\epsilon$ to the results produced by the algorithms, raising once again the tradeoff between privacy and utility. \cite{Dwork:2014vw} states that ``differential privacy requires a new way of interacting with data, in which the analyst accesses data only through a privacy mechanism, and in which accuracy and privacy are improved by minimizing the viewing of intermediate results.'' Differential privacy applies to all subjects equally (due to the specification of a system-wide $\epsilon$); thus it is called \emph{homogeneous privacy}. 

\subsection{Data Science Ethics}
\label{sec:ethics}

The fourth building block of data science is ethics. In many discussions, ethics is bundled with a discussion of data privacy. The two topics certainly have strong relationships, but they should be considered  separate pillars of data science core.

Literature typically referes to ``data ethics'' as ``\ldots the branch of ethics that studies and evaluates moral problems related to data, \ldots algorithms, \ldots and corresponding practices, in order to formulate and support morally good solutions.'' \citep{Floridi:2016}. The definition recognizes the three dimensions of the issue -- data, algorithms and practice -- and provides  a three dimensional design space to talk about these issues. 

\begin{itemize}
\item The \textit{ethics of data} refer to the ethical problems posed by the collection and analysis of large datasets and on issues ranging from the use of big data in a diverse set of applications. The ethical problems related to data include generation, recording, curation, processing, dissemination, sharing, and use. 

\item The \textit{ethics of algorithms} addresses concerns arising from  the increasing complexity and autonomy of algorithms, their fairness, bias, equity, validity, reliability~\citep{Grimm:2021uh}. The relevant problems include analysis algorithms, artificial agents, machine learning, and robots. 

\item The \textit{ethics of practices} addresses the questions concerning the responsibilities and liabilities of people and organizations in charge of data processes, strategies and policies. The ethical problems along this dimension include responsible innovation, programming, hacking, professional codes, user consent, user privacy and secondary use. 
\end{itemize}

The definition is broad in that the term is not used only to cover the ethical issues specific to data, but to ethical issues surrounding the use of the data. This is consistent with the current popular use of the term, but it leads to confusion as we would ideally like to include the ethics issues along all three dimensions. Therefore, it is much better to call this ``data science ethics'' rather than the more common ``data ethics.''

One important issue in data science ethics that has received significant attention is \textit{bias}. Oxford English Dictionary defines bias as ``inclination or prejudice for or against one person or group, especially in a way considered to be unfair a concentration on or interest in one area or subject a systematic distortion of a statistical result due to a factor not allowed for in its derivation.'' Bias is inherent in human activities and decision-making and psychologists have identified three \textit{competing} main reasons for biases in humans~\citep{Baer:2019wk}: \textit{accuracy} for correct decision making, \textit{speed} for fast decision making, and \textit{efficiency} for (perhaps subconsciously) reducing the decision space. Biases in data science generally reflect biases in human decision-making, so it is useful to understand the nature of human biases.

Psychology literature has identified over a hundred types of biases, which are reduced to four for manageability~\citep{Baer:2019wk}.
\begin{enumerate}
\item \textbf{Action-oriented bias}  reflects our common belief that speed is important in many real-life circumstances, so moving fast to make a decision is important. One type of action-oriented bias is \textit{bizarrness effect}, which is the tendency to highlight something that is unusual and brand all  things that are similar to it as being the same. \textit{van Restorff effect} is also an action-oriented bias where one focuses on a single thing that stands out to the elimination of other possibilities. An action-oriented bias that shows up considerably in data science is \textit{overconfidence} where the person is so certain of his views that he does not consider alternative solutions, approaches or interpretations. Note that all of these biases reduce the alternatives that need to be considered, and therefore reduce the decision-making time.

\item \textbf{Stability bias} ``refers to the human tendency to act as though one’s memory will remain stable in the future''~\citep{Kornell:2012vp}. These biases represent a preference for the status quo and a reluctance to change. An important stability bias that shows up in data science is \textit{anchoring effect}, where one's current mental positioning determines where she goes (or willing to go) next.

\item \textbf{Pattern recognition bias} allows people to recognize patterns to fill in gaps in their understanding. One very common type is \textit{educated guess} that allows people to use their partial knowledge and reasoning to draw conclusions. In other words, they allow filling in the blanks (in data) -- usually following ``better safe than sorry'' principle. When these guesses are systematically wrong, they lead to pattern recognition bias. A particularly challenging type that also appears frequently in data science is \textit{confirmation bias}, which is the  ``tendency to search and select information confirming personal hypotheses and beliefs, ignoring contrary evidence''~\citep{seel:2012vu}.

\item \textbf{Interest bias} answers the question ``What do I want'' and differs from the others whose objective is to make correct decisions as accurately, quickly, and efficiently as possible. It reflects human belief that what we want is generally good for others. A particular exhibition of this type of bias is \textit{social bias} which is reflected as groupthink or going along to avoid harm to one's self even when it is known that the decision is wrong.
\end{enumerate}

Bias is inherent in human activities and decision-making, and human biases are reflected in data science as \textit{biases in data} and \textit{biases in algorithms}. Bias in data can be introduced through what is included in the historical data used by the algorithms. For example, the arrest records in the U.S. include more marginalized communities, primarily because they are over-policed. Algorithms are employed thgat use this data for decisions and recommendations. Bias in data may also be introduced due to under-representation. For example, data used in face recognition systems contain 80\%  whites of which three-quarters are males. 

Four types of data bias can be identified ~\citep{Srinivasan:2021ub,Krishnamurthy:2019tc}:

\begin{enumerate}
\item \textbf{Collection bias} consists of errors that are made in collecting data. Some data is collected from user surveys and reviews and these can be supplied by a small number of people whose views get represented far more than their occurrence in the population. For example, less than 10\% of the registered users generate 50\% of the reviews on Amazon and Facebook~\citep{Baeza-Yates:2018vk}. Even the presentation of data (e.g., its order) might have an impact on the responses one gets. Societal bias also plays an important role in collection bias as respondents' answers will be influenced by the biases of the societies in which they live and operate. Finally, measurement bias involves the imprecision in collecting data or the effect of confirmation bias as to what data is collected.

\item \textbf{Selection bias},  also called sampling bias, occurs when the dataset used in analysis differs considerably from the real data. It arises in a dataset that is created by selecting particular types of instances more than others. Selection bias can be the result of poor planning or the result of confirmation bias; it can even be due to laziness as the data that is easiest to find is picked. A common example of selection bias is the lack of proper gender representation in the datasets~\citep{Perez:2019vk}.

\item \textbf{Stability bias} is the result of not taking into consideration possible changes in data. It is also called historical or drift bias. In many cases, the analysis is conducted on static data with no accommodation for possible changes. However, there could be periodic (e.g., seasonal) changes or changes in the data distribution due to changes of the underlying phenomenon, and if these are not captured, the model will perform poorly when data changes. 

\item \textbf{Aggregation bias} occurs when data is aggregated, resulting in loss of granularity and fidelity. For example, averages hide  significant information -- outliers are lost in averages, but they can also push the average in a direction that does not properly reflect the data distribution.
\end{enumerate}

The second class of biases are those in algorithms. Algorithmic bias can occur due to the inclusion or omission of features in the algorithm/model\footnote{In machine learning deployments, it is common to talk about models rather than algorithms; for the purposes of this paper, they are similar in that they capture the behaviour of the underlying system or phenomenon. Therefore, the terms are used interchangeably.}. In machine learning deployments, this can occur during feature engineering. These features include individual attributes such as race, religion, national origin, gender, marital status, age, and socioeconomic status. Sometimes one of these attributes is omitted with the claim that the algorithm is free from bias related to this attribute. This may indeed be true in some cases, but is not generally true. For example, omitting race as an attribute does not automatically make an algorithm non-racist since, in some locations, postal codes are correlated with race and can reveal race information. Similarly, omitting gender from the set of considered features does not make the algorithm gender-agnostic. 

The use of proxy metrics in place of metrics that are not easily measurable can also lead to bias.  For example, in graduate student admissions, the decision problem is whether an applicant would be successful in the graduate program, and standardized examinations are used as proxies to judge future success in the program. However, the unreliability of these tests in predicting student success has been well-established. 

Four types of algorithmic bias can be identified:  

\begin{itemize}
\item \textbf{Feature bias} is the most common type and is introduced when decisions are made as to which features to include in the algorithm; if a feature is not included, generally its effect may not be seen unless there are other correlated features from which deductions can be made. Missing features can be seen as an example of overconfidence bias (I know what I am doing and that is not important) or confirmation bias (I know that this feature is the most important).

\item \textbf{Confirmation bias} shows up in model building/algorithm design because most often we need to define what the dependent variable (in machine learning this is called the target function) means. We typically define one result as good and another result as bad. For example, in banking the decision is to whether or not give a loan to an applicant. Confirmation biases interfere with the definition of ``good'' and ``bad'' for the dependent variable.

\item \textbf{Algorithm logic bias} occurs when the algorithm is designed to be biased. A well-documented example is the use, by some healthcare providers in the U.S., of an algorithm in determining patients who should get additional health care assistance. The algorithm has been determined to be racially biased, because it focuses on predicting health care costs rather than illness, ``but unequal access to care means that we spend less money caring for Black patients than for White patients''~\citep{Obermeyer:2019vo}. Another well-publicized example is the algorithm Amazon used in recruiting that showed bias against women and was eventually dropped.

\item \textbf{Overfitting/Underfitting} represent algorithmic errors where the algorithm has a particular accuracy on the training data, but a different one on the unseen data to which it is applied. In the case of overfitting, the model fits the training data too closely and cannot generalize. Therefore, it will not perform accurately on the unseen data. This can happen if the model is too complex or the training is too long. Underfitting is the opposite -- the model does not accurately capture the behaviour of the training data. This can happen if the training duration is too short or if the model uses features that do not capture the relationship between the independent (input) and dependent (output) variables. In either case, the result is bad predictions.

\end{itemize}

Although considerable attention has focused on the problem of bias, data science ethics should be considered more generally, consistent with the previously offered broader definition of ethics. Some of the broader ethical considerations have overlapping concerns with data protection. Furthermore, the ethical concerns reflect societal norms that usually get coded  in legislation. Therefore, while some of the ethical concerns are universal, others can be specific to a particular jurisdiction.  The broader ethical concerns include \textit{ownership} considerations that address who has ownership of data, \textit{transparency} that refers to subjects knowing the data that is collected about them, how it will be stored and processed (including informed consent by the subject), \textit{privacy} of personal data, in particular the revealing of personal identifiable information, and finally, \textit{intention} regarding how the data will be used, especially for secondary use. As noted, there are jurisdictional differences in how these concerns are addressed. For example,  North America and Europe have different viewpoints on ownership, where the latter is more attentive to individuals' ownership of personal data. 

Berkeley Data Science Institute has develop a checklist that can be useful in discussing data science ethics~\citep{Lou:2020ws}. The checklist contains the following questions:

\begin{itemize}
\item Have we listed how this technology can be attacked or abused? 
\item Have we tested our training data to ensure it is fair and representative? 
\item Have we studied and understood possible sources of bias in our data? 
\item Does our team reflect diversity of opinions, backgrounds, and kinds of thought? 
\item What kind of user consent do we need to collect to use the data? 
\item Do we have a mechanism for gathering consent from users? 
\item Have we explained clearly what users are consenting to? 
\item Do we have a mechanism for redress if people are harmed by the results? 
\item Can we shut down this software in production if it is behaving badly?
\item Have we tested for fairness with respect to different user groups? 	
\item Have we tested for disparate error rates among different user groups? 
\item Do we test and monitor for model drift to ensure our software remains fair over time?
\item Do we have a plan to protect and secure user data? 
\end{itemize}

\section{Data Science Applications -- Examples}
\label{sec:apps}

As noted earlier, applications are crucial in data science as they give purpose to the entire endeavor and serve to inform the core technologies. A question that is commonly asked is what constitutes a data science application. The short answer is that any domain that has and works with very large datasets is a good candidate for the application of data science principles, techniques and methodologies.

In this section, the objective is to go beyond this simple statement and discuss some example applications. The point is not to be exhaustive in the set of applications or complete in the discussion, but to present some examplars.

\subsection{Sustainability}
\improvement{Citations needed in this section.}

The world is facing multiple, intersecting, and systemic challenges. Global environmental problems are
reaching crisis points including climate change and biodiversity loss, but also freshwater depletion and
pollution. Meanwhile, we are off track with meeting our numerous human development objectives such
as reducing poverty and hunger, as shown by trends in indicators of the Sustainable Development Goals
(SDGs). Navigating pathways to a sustainability transition is a major challenge faced by the scientific
and policy community, and society at large. 

At the same time, information about our world has never been more abundantly and easily available.
The availability of data from many sources is increasing, such as earth observation datasets and data
analytical tools, satellite-based remote sensing data, ground-based observational data, and citizen science,
high spatial and temporal resolution data from digital technologies such as cell phones and
drones. As a result of the growth of massive multi-dimensional data sources (petabyte scales in some instances), new approaches are needed to facilitate advanced analysis of complex natural, hydrological and atmospheric science systems, as well as human societal systems. Advanced geospatial analysis, the emergence of open software communities, and the development of open data repositories that include data regarding private entities and about people’s lives (e.g., health, environment, economic data, crime and safety, census demographics) are essential to solving pressing questions related to climate change, ecosystem services, and human decision-making. Data science is expected to play a key role in sustainability transformation -- by harnessing the large amounts of environmental and social data that are becoming increasingly available, data science can be used to catalyze change that transcends disciplinary boundaries~\citep{Dunn:2021wh}.

The state of the environment has long been scrutinized and investigated through the analysis of data from in situ observation or survey-based techniques. Typically, this data forms the basis for explaining social and natural environmental processes; the idea that the state of an environmental system can be diagnosed by direct ``measurement'' is a classic proposition that has underpinned environmental science for decades. Direct measurement encompasses a broad scope of technologies and applications that can be used not only to observe a system from multiple perspectives, but also to drive complex social and natural environmental system models.

\paragraph{Climate Variability and Change.} The availability of satellite observations and model simulations, along with in situ measurements of the state of Earth’s systems, underpins our ability to quantify, characterize, and contextualize changes within the physical environment. However, the proliferation in the availability of datasets is significant to the extent that it is now nearly impossible for researchers studying environment and climate change to conduct analysis of trends, across inter-related datasets at very large spatial scales, and over decadal time scales, without the use of data science. Consequently, the generation of continental-scale, petabyte-sized multi-dimensional data models enables researchers to exploit systematically-derived, analysis-ready datasets that are by their very nature, designed to address these issues.

Using complex analysis methods (e.g., geostatistical analysis, object-based analysis, machine learning), these multidimensional models can be utilized to address a multitude of environmental problems; including use/land cover changes that impact flood inundation; water resources management related to changing snow and glacier ice mass; aboveground biomass (e.g., forest) states and changes that impact carbon sinks, as well as sources that are affected by fire and destructive invasive species (e.g., mountain pine beetle). Tangibly, satellites have provided consistent observations of Earth’s systems for the past 40 years, while climate model simulation experiments can extend our knowledge to the pre-satellite era. Additionally, in situ environmental data from operational networks also extend available data to the early and mid 20th century.
 
However, the analysis and fusion of these datasets to derive as much information as possible about environmental and climate change is currently only feasible at small-scales (limiting our collective understanding of the problems we study) without fundamental approaches to data preparation and management. Furthermore, the amount of data available for exploitation is only set to grow in the future, placing more pressure on the existing data science technologies.

\paragraph{Ecology.} There are major challenges to ensuring effective management, conservation, and restoration of ecozones (i.e., large biogeographical regions usually defined by major divisions of bedrock or climatic isoclines). Aside from the issues of addressing ahistorical dynamics caused by anthropogenic climate change and land use change, there are issues with data scales. For example, data is collected on small (metre-scale plots) and large scales (decadal time-series covering ecological landscapes that include watersheds or an entire ecozone of 100,000 km$^2$). Large-scale datasets  are, by definition, big data, but so too are the cumulative datasets from smaller scales that add up quickly to become distributed datasets. 

To ensure that biodiversity and ecosystem functions (and services for humans) are going to maintain resilience in the face of climate change and land use like urbanization or resource extraction, there is an urgent need and opportunity to take advantage of these underutilized datasets, which are currently incompatible by the lack of appropriate data science infrastructure. The drive within the ecological sciences domain is to use diverse sets of big data, where effective data preparation and management, improved analysis through the application of new tools, and advanced methods to visualization can lead to novel insights, robust results, and increased explainability for end-users. End-users range from individual ecosystem managers through to government policy-makers, all of whom would benefit from the ability to fully exploit this previously inaccessible data. New data can be processed in real-time, leading to faster and more appropriate decisions about ecosystem management. The objectives of this line of research extend to modelling of urban flood risks, spatial optimization of restoring old resource extraction sites (e.g., oil wells, gas wells, quarries, mines), and modelling corridors at landscape scales to connect natural areas and make them more effective at conserving biodiversity.

Various ecology research clusters are responding to the natural and anthropogenic ecosystem changes that are increasing in rate and severity. Data science can provide the critical infrastructure to translate long-term monitoring and assessment into solutions. We are moving beyond current capabilities by assembling largely portable infrastructure that will add integration of ground-based data collection with larger data-streams and new approaches to measuring ecosystem change using unmanned aerial vehicles (UAVs). This increases the volume and the variability of the datasets that are used in analysis.

\subsection{Power and Energy}

A key feature of the electricity sector is the complexity of its interactions amongst several layers of systems, physical assets, and data flows. Thus, not one specific data tool is needed; a combination of multiple datasets and techniques are required to fully leverage all available resources to achieve the desired outcomes~\citep{Zhang:2018wn}. The electricity market is  being  restructured to accommodate distributed generation technologies as part of a movement towards clean energy solutions. One of the biggest advantages of employing data science in support of energy management is the optimization of energy production and distribution~\citep{Zhang:2018wn}. Power generation from clean energy technologies is on the rise, but the intermittent and unpredictable nature of these resources (for example, sunlight and wind sources) hampers overall energy production. As a result, it becomes difficult for solar and wind power plants to operate at their maximum potential. Data science is rapidly changing this scenario. Through predictive analysis, valuable data is now combined with weather and satellite equipment, which leads to highly accurate forecasting of weather conditions in advance~\citep{Sweeney:2020wl}. This allows renewable energy plants to optimize their production significantly. Data analytics can help companies to produce more energy without requiring additional infrastructure costs. This, coupled with the increasing optimization of power grids, is a contributor to the steady decline of renewable energy prices. In the near future, renewable energy will be comparable in cost with its conventional counterparts.

\paragraph{Power Systems.} Power systems are rapidly evolving into digital systems through the deployment of Smart Grid infrastructure~\citep{Farhangi:2010ui,IESO:2011wf} -- also see, as an example, ~\citep{IESO:2011wf}. This new infrastructure allows for two-way communication between utility companies and their customers, and sensing technologies along transmission lines makes the grid “smart”. With this new technology, the grid can respond in real-time to changing electricity demands, making transmission more efficient, lowering costs for operations and for consumers, and allows for better integration of new renewable energy systems. Electric utilities worldwide are embracing the Smart Grid vision, which includes full modernization and automation of electric power networks. A Smart Grid approach seeks to bring together and connect, in an interoperable way, diverse technologies: advanced applications and use of distributed energy resources, communications, information management, and automation. In doing so, the Smart Grid enables a safe, self-healing, reliable, less constrained, and more efficient grid. Intelligent meters, products, and displays also empower customers to use electricity more efficiently.

The entire energy supply chain can tremendously benefit from the application of data science technologies, as they can improve the optimization processes in power grids, and enable the effective management of large amounts of data~\citep{Tu:2017wb}. Over the past decade, the accumulation of relevant data has accelerated. The advancement of data science techniques in power systems can lead to real-time optimization of electricity generation and transmission, prediction of load supply and demand, consumption pattern analysis that can lead to new services, and dynamic pricing strategies. 

A data science approach will enable a collaborative effort in finding solutions to the challenges arising in the operation of Smart Grids, not from the individual viewpoints of power systems, communications, or information management, but through a holistic approach~\citep{Mazzeo:2021aa,Li:2019vu}. For example, a smart home energy management device needs communication channels with individual appliances, but it cannot stand alone – it needs information from the external power system in terms of real-time market prices. Therefore, every activity in a Smart Grid system must be linked between three important pillars: the power system, the communication system, and the information system~\citep{Mazzeo:2021aa}. 

\paragraph{Renewable Energy Management.} An increased emphasis on predictive analytics is necessary to manage grids based entirely on renewable energy due to the variable nature of the energy sources. Another factor to consider is that with renewable energy, consumers may also have the ability to contribute energy back to the grid. In order to tackle increasing energy demands, solutions for efficient energy consumption, generation, and distribution need to be assessed through the lens of data modelling and management. Data science, in combination with other information and communication technologies, is key to empowering modern electric grids with the ability to support two-way energy and information flow, expediting the integration of renewable energy into the grid and providing the consumer with tools for optimizing energy consumption and costs. The integration of ICT and smart data into traditional electricity infrastructure and clean energy technologies, such as solar, wind, geothermal, biomass, and storage, will lead to the development and implementation of efficient Smart Grid networks and architectures~\citep{Canizares:2019vk}.

\subsection{Biological and Biomedical Systems}

The interdisciplinary fields of bioinformatics, computational biology, and mathematical medicine include a number of problems that are highly dependent on data and would benefit from the application of data science techniques and methods~\citep{Supriya2021}.

\paragraph{Bioinformatics.} Data science techniques are at the heart of many, if not most, of the computational methods used in bioinformatics, including its subfields of genomics, transcriptomics, and proteomics~\citep{Luscombe:2001vu,Gauthier:2019ws,Hagen:2000vc}. For this reason, the field of bioinformatics and its subfields have been described as “biological data analytics” or as “biological data science”.
The boom in biological data science over the last decade is largely due to recent advances in DNA sequencing technology and the field of genomics. It has become routine to sequence not only whole genomes, but also populations of genomes and “metagenomes”, containing DNA from thousands of microbial organisms from environmental samples. This explosion of sequence information is responsible for driving an enormous wave of research within bioinformatics, yielding new methods for assembling genomes, identifying the location of genetic elements, and predicting their biological function. 

The exponential growth of sequence data has faced challenges in the management, accessibility, and retention in using, storing, and sharing these datasets~\citep{Benson:2013de,Hunter:2014tv}. The explosion of genomic data has been paralleled by the development of technologies for generating different types of high-throughput biological data, such as transcriptomics, proteomics, metabolomics, lipidomics, and atomic-resolution three-dimensional structures of biological macromolecules (structural biology). Each of these data types requires specific techniques for data preparation and management, facilitating large-scale data analysis, and interpretation and explanation of results. 

\paragraph{Genomics.} Due to advances in high-throughput DNA sequencing, the costs of genome sequencing have plummeted, leading to a substantial increase in the amount of sequence data processed by research labs and biological databases. In a typical workflow, millions of raw sequencing reads are collected that represent short, overlapping, and redundant fragments of DNA from either a single-species source (i.e., whole genome sequencing) or a community of microorganisms associated with an environmental sample (i.e., metagenomics). %Metagenomics commonly includes “microbiome” sequencing of host-associated samples, such as the human gut, or environmental DNA sequencing, from virtually any habitat containing life (e.g., aquatic biomes, soil or terrestrial habitats, engineered environments).

There are key data science concerns for the task of whole genome sequencing of a single species: genome assembly, genome annotation, and comparative genomics and biodiversity informatics. Genome assembly requires graph theory-based analysis to reconstruct large genome sequences from short, fragmented, but overlapping reads~\citep{Miller:2010vb}. The challenge concerns the management and analysis of these growing datasets, which increasingly require processing and analyzing graph-structured data at scale. %Genome annotation involves finding the locations of genetic elements (e.g., genes and regulatory regions) within an unannotated genome sequence. Hidden Markov models (HMMs) can be used to identify sequence matches to predefined gene or protein family models; however, these methods do not scale well with large datasets. Instead, this problem may be addressed using hashing-based techniques for rapidly assigning functions based on precomputed database annotations. 
Comparative genomics and biodiversity informatics typically focus on identifying genomes that are closely related through sequence and gene comparison, chromosome structure, or raw sequence. These comparisons require many data science techniques, including digital signal processing, combined with alignment-free comparison techniques, or hierarchical clustering for building trees, set comparison, and sequence alignment.

%Sequence variant detection (i.e., mutations, single-nucleotide polymorphism, or SNPs) is also an area of interest, particularly the analysis of these variants to identify those that may be responsible for disease or phenotype. These analyses are commonplace in population sequencing initiatives or GWAS (genome-wide association studies) analysis of human disease. For this task, several fundamental data science approaches are needed, including data preparation, management, and analysis. Explainability is also a key challenge to ensure that end-users can interpret the insights produced by the analysis; for example, in the classification of disease based on SNP profiles.

%Metagenomics is the study of genomes recovered from environmental samples, which may contain genomes from multiple organisms or individuals; this is different from genomics, which is the study of the complete genetic information of one organism. For metagenomics, the data science concerns are similar to those noted for genomics, plus several domain-specific problems, such as taxonomic profiling (estimating species populations present within a sample and their relative abundances), and analysis techniques for identifying data and metadata associations. The latter includes applications like prediction of gastrointestinal disease states given gut microbiome data. For example, if a taxonomic profile (frequency) contains elevated levels of certain pathogenic bacteria, this may be diagnostic or predictive of gut inflammation or other diseases.

\paragraph{Transcriptomics.} Transcriptomics uses sequencing or microarray technologies to measure the levels at which all transcribed genes within a sample (e.g., cell type, tissue, or whole organism) are expressed~\citep{Wang:2009ur}. Transcriptomics has enabled the study of how gene expression changes in different organisms and has been instrumental in the understanding of human disease. The typical transcriptomics analysis workflow involves quantification of transcript abundance, clustering of samples based on overall transcriptomes, and statistical comparison between samples to identify “differentially expressed” genes. Data science techniques are essential for these tasks. For example, principal components analysis (PCA) is used to organize samples spatially based on their gene expression profiles. Since all genes are individually analyzed to compare their abundances among samples, this necessitates the use of sensitive and high-throughput statistical techniques for p-value determination, such as the calculation of false discovery rates. 

\paragraph{Proteomics.} Proteomics is the large-scale study of all proteins in a cell, tissue, or organism~\citep{Aebersold:2003tb}. The proteome is not constant, but instead differs among cells and tissues and changes over time and with environmental conditions. It can provide a snapshot of a functional biological system, offering a detailed view of the intracellular state; it can also be used to investigate where and when proteins are expressed, what functional modifications to proteins are present, how they are involved in metabolic pathways, and how proteins interact with cellular substrates and with one another. For example, proteome and genome information can be used to identify proteins associated with a disease, which algorithms can then use as targets for new drugs.

Data science methods are crucial for processing proteomics data. The assignment of peptide sequences to mass spectrometry molecular fragmentation patterns requires extensive analysis via pattern recognition techniques and estimation of false discovery rates. Quantification of peptides and proteins requires in-depth statistical analysis and corrections for multiple hypothesis testing. Mapping these data onto biochemical pathways can provide insights into the functioning of the biological system, requiring that analysis techniques are explainable and interpretable to proteomics experts. Collaboration with data science experts will ensure that methods will yield accurate and robust results that can be effectively translated to the clinical setting.

\paragraph{Computational Systems Biology.} The three different -omics fields of bioinformatics described above can be integrated to form a combined multi-level perspective of a biological system~\citep{Cusick:2005wh}. This is the major goal of systems biology, which combines diverse biological data types into computational models and predictive frameworks. For the analysis of systems biology data, a common and powerful technique is the combination of graph-based approaches and Bayesian statistics for combining different evidence types to predict functional interactions between different components of a biological network. In this way, multi-omics data can be analyzed to produce “interactomes” that aim to characterize the molecular interaction networks within cells and explore their dynamics. Future work in this area will require integration between data types (requiring data preparation and management techniques) and interdisciplinary interactions between bioinformatics and data science researchers with expertise in different areas.

\paragraph{Mathematical and Computational Medicine.} In a biomedical context, there is information transfer over multiple scales, ranging from the genome, through the proteome, to tissue level manifestations of disease. Similarly, it has become clear that there is also a reverse cascade, by which mechanical forces at the tissue or organ level can impact the genome (through mechano-transduction and other mechanisms). The purpose of mathematical medicine is to develop multi-scale models (for example, macroscopic continuum models coupled to cellular level models) to allow transfer of genetic and cellular level information that might permit development of appropriate boundary and initial conditions for the continuum level model. For this purpose, not only are the theoretical tools from partial differential equations and dynamical systems of paramount importance, but the emerging area of data-driven dynamic systems will also play a central role~\citep{Albers:2018ts,Baker:2018vk}. The combination and integration of these approaches with new analysis techniques heralds the prospect of dramatic advances in our understanding and treatment of major problems, such as drug-induced drug resistance, optimization of combination therapies, and the design of patient-specific “personalized” medicine. Therefore, data science is needed to ensure optimal combination of data types through appropriate data preparation, as well as accurate analysis. As the results will provide insights for medical decision-making, it is critical that explainability and interpretability are incorporated into the research.

\subsection{Health Sciences and Health Informatics}

The explosion of health data and the growth of associated analysis techniques are delivering substantial and rapidly growing benefits to research and practice in health and healthcare. Traditional sources of data in these fields have been growing in both size and diversity. They include administrative and clinical assessment data, large-scale survey instruments, public health records (e.g., reports of notifiable illness, including dynamics in infectious diseases), laboratory and imaging results, and electronic medical records. Increasingly, these sources are supplemented by a rich set of electronic data sources, such as social media, data from wearables and mobile devices, electronic health records, communication behavior, and search data. The trend in the field has been to manually cross-link multiple data sources for stronger insights. Additionally, while traditional biostatistical methods remain of key importance, these datasets are increasingly analyzed with machine learning methods and causal statistical models.  In some areas, the results have been transformative. Research in the health sciences has been combining these growing diverse datasets, and advancing the types of analysis used, in order to provide holistic insights into healthcare challenges; the effective combination of diverse data types requires knowledge of data preparation and management to ensure robust results~\citep{Consoli:2019uh}. 

For example, social media and search data have begun to provide novel means of syndromic surveillance and insights into the dynamics of knowledge, attitudes, and beliefs. Data from mobile technologies and wearables have enabled the use of physical measures to study key health behaviors and exposures that were previously infeasible due to costs or participant burdens to collect this data. Now, researchers can collect data like contact and mobility patterns, diet, physical activity, sedentary behaviour, and communicational behaviour. A growing body of work suggests that use of such data offers the potential to provide not merely a more detailed view of well-understood behaviors and risks, but can, in some cases, dramatically alter study research findings. Such data have supported recognition of common biomedical drivers of health outcomes, resolution of health microbehaviors, enabled quantification of dose of exposure to certain environments (e.g., poorly walkable or unsafe neighbourhoods) and influences (e.g., tobacco billboards), and most importantly, shifting the patient engagement model from a discrete (single clinical visits) to a more continuous view of patients’ health.
 
Such electronic data has also supported the recognition of symptoms and supported identification of latent health states (e.g., diabetes or depression), and of effect (e.g., stress levels). There is a notable and growing interest among health scientists and practitioners in data-intensive approaches to inform insights in health via analysis of large-scale datasets. Data-intensive health sciences include public health surveillance, health services research, environmental surveillance, global health, health informatics, precision medicine, mobile health technologies for chronic disease management, continuing care, aging, mental health, intensive care unit patient management, and healthcare system usage.

Such efforts can provide guidance at many levels, from clinical to health service management to public health. While the growing use of dynamic modeling in health policy-making has seen little impact by data science advances, lines of recent work have successfully secured benefit by combining dynamic modeling with high-velocity evidence for informed decision-making, such as with sequential Monte Carlo methods~\citep{Li:2018wr}, deep learning approaches~\citep{Watson:2021tl}, and dynamic prediction approaches~\citep{Jarrett:2020tv}. The use of natural language processing (NLP), based on sequential deep learning approaches (e.g., Long Short Term Memory (LSTM) and recurrent neural networks (RNNs)), has shown great promise for sifting through and discovering patterns and trends in the vast array of data that is generated online by social networks, both private and public.
 
The growing pervasiveness of sensing devices and online connectivity in the homes of people living with health conditions (e.g., smart-homes for people with age- or disability-related conditions) has led to the generation of massive dynamic datasets that can be leveraged to improve health outcomes. Through judicious use of real-time analytics techniques, it is possible to produce more effective solutions in care. For example, the detection of wandering or falls in the elderly can lead to massive time savings for human caregivers, reducing the need for continual monitoring.  

Despite the progress and potential noted above, there remain far more untapped potential for data science in health; a few are described below.

\paragraph{Data Privacy and Management.} Perhaps the most important aspect of leveraging data for research in healthcare is the overwhelming importance of privacy in health data, and correspondingly, the need to operate within the strictures of the existing legislation governs much of this data, e.g., Canadian Personal Health Information Privacy Protection Act (PHIPPA) and the privacy provisions of U.S. Health Insurance Portability and Accountability Act (HIPAA). Despite the cross-sectoral nature of many health challenges (e.g., mitigating the opioid epidemic or drug resistance) and profound attendant interest in cross-linking data across jurisdictions, health scientists are currently facing the reality of needing to operate with balkanized silos of data. Where cross-linking does occur, it is typically performed in a manual and ad-hoc fashion on the raw data, with the linked data then stored at one or more locations, with attendant risk to privacy. Analyses are then typically also performed manually, with the same privacy considerations.

\paragraph{Data Accessibility and Explainability.} In many jurisdictions, patients (and their families) are typically provided little or no access to their own data, thereby limiting prospects of (assisted) self-management on which sustainability of the healthcare system increasingly depends. When seeking to conduct analyses of large-scale data, health scientists typically remain almost entirely dependent upon computer scientists for generating, explaining, and visualizing the results -- this greatly crimps learning, and almost entirely prevents the routine application of these results by health decision makers.  While recent contributions have demonstrated great potential for conducting broad-scale syndromic surveillance with electronic data feeds, the fragmentation and shifting nature of the interfaces, as well as the technological challenges in realizing these mechanisms, place widespread use of these technologies out of reach for health agencies. This leads to vastly varying privacy terms that both confuse health research ethics boards and pose serious barriers to the effective institutionalization of such surveillance.

\paragraph{Data Analysis.} Existing systems and platforms in this domain fail to provide scalable implementations of core machine learning and statistical algorithms (e.g., forms of time-series analysis such as LSTMs and RNNS, and forms of time-to-event analysis~\citep{Jarrett:2020aa}); therefore, powerful combinations of dynamic models remain so cumbersome that they are still inaccessible for many applications in the health and healthcare fields. The necessity of real-time analytics, as noted above, is also a substantial challenge.

Advances in privacy-preserving data mining and homomorphic encryption could provide avenues for the advancement of this research while enhancing public and regulatory confidence. The use of blockchain and other technologies promise means for securing flexible distributed storage of data in a manner that simultaneously provides the patient (and family) access to their information for decision-making, and assistive technologies can be used to provide fine-grained dynamic gating of access to both cross-sectional and longitudinal personal data.  Collectively, advancement of these privacy-enhancing technologies raises prospects for the use of data science techniques on richer cross-linked data than is now feasible. Employing data science techniques in the development of these technologies will yield particular attraction within the health and healthcare domains, where privacy is paramount and clear guarantees of privacy can be enabling in terms of securing buy-in for widespread use. 

\subsection{Digital Humanities}

The study of human culture and activities has been reinvigorated by the growth of available data and digital tools. Cultural information, either that has been digitized from earlier material (e.g., a 19th century newspaper that has been scanned, indexed, and made accessible) or that begins its life as a digital object (e.g., social media streams or statistical information), is increasingly accessible to humanists and social scientists~\citep{Milligan:2019tu,Romein:2020vc}. Similarly, scholars now have the ability to present their findings in increasingly innovative ways – interactive visualizations, webpages, databases, 3D environments – and to engage with the broader public in new ways. These trends have given rise to the field known as the digital humanities. A broad term for a wide range of scholarly activity, the digital humanities are primarily, but not exclusively, concerned with both understanding how technology is impacting and changing the study of human culture and activity, as well as the changing modes of scholarly communication and impact. 

These are aligned with objectives in data science. Humanists have been engaging with these digital questions for decades, but are increasingly recognizing the necessity of interdisciplinary collaboration. 

\paragraph{Digitization of Human Culture.} Recent progress in the digital humanities has focussed on transforming scholarship through the digitization of human culture and activities, as well as the methods for dissemination of this scholarship. While the humanities have been traditionally defined by print publications, often solo-authored, the digital humanities challenges this through emerging technologies. For example, new tools are being developed to digitize primary documents (not only text, but sound and image as well) to provide immersive historical experiences to disparate audiences, including students, the general public, scholars, and policy-makers. An overarching area of research is the exploration of new methods and technologies for open, interactive, and collaborative argumentation to bolster the case for humanistic inquiry, that is, the act of searching out a general understanding about the world around us~\citep{Robertson:2021ul}.

As humanists digitize data and think of new ways to present it to diverse audiences, they suddenly find themselves needing considerable data science skills: from data preparation (how to map analog documents over to modern databases), to management (some data may be open to all; other data may be restricted to particular communities; and how to keep it sustainable so that future scholars can benefit), to analysis (applying new methods to existing challenges and data), explanation (how to make datasets intelligible and reusable), to dissemination (how to share datasets and insights). As more projects migrate to an online format, they are increasingly discussing new frontiers in linked open data to break silos down between repositories of digitized knowledge. Beyond data scientists, this interdisciplinary field can see collaboration between historians, other social scientists and humanists, as well as policy-makers, activists, and beyond\footnote{See, for example, LINCS project: \url{https://lincsproject.ca}}. Making data science accessible to humanists also heralds considerable pedagogical value.

\paragraph{Impact on Humanistic Inquiry.} Another area grapples with the impact that new digital environments and worlds are having on humanistic inquiry. The Internet, video games, virtual reality, and other new and emerging platforms and formats all present new frontiers for humanistic inquiry: what does it mean to ``play''? What can we learn from serious video games~\citep{Chapman:2017vr}? How can we understand the Internet as a platform? What are the ethical implications of working with the large-scale data that is generated on these platforms~\citep{Lomborg:2013vl}? Similarly, there is research that explores how historians can use born-digital resources, such as archived websites and social media streams, as well as the types of digital infrastructure~\citep{Milligan:2019tu} – and ethical apparatuses – that will be necessary in the digital age.

Here, too, as humanists begin to grapple with data in diverse formats (from video games to archived web pages to virtual environments), data science begins to intersect in new and novel ways. This data must be managed, especially as it begins to reach into the scales envisioned, raising questions around long-term preservation, retention schedules, and beyond. Experience matters too: how does a video game emulated on a modern system compare to an earlier one; or how is a 1990s website presented on a modern LCD monitor? Interoperability becomes increasingly important as well, as humanists generate and work with data at scale, they need new forms of dissemination and explanation\footnote{See, for example, \url{https://rhizome.org/editorial/2014/oct/22/big-data-little-narration/}}. These will raise fruitful moments of intersection between digital humanists and data science. 

\subsection{Finance and Insurance}

The finance and insurance industries aim to discover value and manage risk. The increasing availability of data is changing business focus and operations in these industries~\citep{Chakravaram:2019tg}. A few current research themes of data science in finance and insurance include risk analysis and management, financial modelling and decision-making, and financial technology (FinTech). Availability of large and varied data (i.e., big data), contributed through social media and a large number of transactions, has revolutionized the way in which financial institutions operate. Data in these fields is present in both structured and unstructured formats. This data can be leveraged for several applications in finance and insurance, as described below.

\paragraph{Risk Analysis and Management.} Risk analysis and management is a key facet of the finance and insurance industries, with risks including competitors, credits, and markets, to name a few. The main steps towards managing risks is to identify, monitor, and prioritize. As with many applications, the amount of data available for analysis is growing substantially, and includes variables like details on financial transactions and consumer information. Data analysis is used by financial institutions to process this data to verify the creditworthiness of customers, compute risk scoring models, and to optimize their costs. Ultimately, this information is used to make strategic decisions~\citep{Buehler:2019vz}, and to increase trustworthiness and security. The management and analysis of these growing datasets, as well as the security and privacy concerns of this sensitive data are key concerns for the advancement of these technologies. Additionally, as the insights produced by these approaches are used to inform decision-making, the explainability of these models is critical to ensuring the transparent dissemination of these results to end-users at these institutions. 

\paragraph{Consumer Analysis.} Consumer personalization is a major focus of financial institutions. With the help of real-time data analysis, data scientists are able to take insights from consumer behavior and make suggestions for appropriate business decisions. Financial institutions, such as insurance companies, conduct analyses to measure the customer lifetime value, to increase their cross-sales, as well as to reduce below zero customers for optimizing losses. Furthermore, financial institutions are also relying on speech recognition- and natural language processing-based software to provide better interactivity with their users. With the data that is provided by users, financial institutions are able to take actionable insights based on their customer needs, which can lead to an increase in profit. This helps institutions to optimize their strategies and provide better services to their customers.

\paragraph{Valuing and Hedging Expensive Portfolios.} Model evaluation of many complex portfolios (e.g., portfolio of variable annuities) held by insurance companies or banks involves double-nested Monte Carlo computations that are so expensive that they are currently not properly evaluated or hedged~\citep{Broadie:2015wj}. Machine learning approaches (e.g., neural network or regression-tree-based methods) can significantly reduce the compute time while maintaining a reasonable accuracy. Leveraging fundamental data management and analysis tools is a promising approach for the effective management of expensive financial and insurance portfolios.

\paragraph{Economic and Financial Instruments for Climate Change.} Global climate change, environment, and pollution have recently led to substantial economic and financial costs. In addition to the traditional approach of command-and-control measures (e.g., taxes to reduce pollution), studies have shown that economic instruments, such as emission trading, can be a more flexible and cost effective alternative. For instance, a number of jurisdictions have introduced a cap and trade system for the electricity sector  to reduce air pollutant. Emission permit trading is a policy tool that is used to reduce carbon emissions. Emission targets should be set in a way that is cognizant of the trade-off between immediate cost of emission reduction policies and the long term risk of catastrophic climate change. Data science approaches to model the trade-offs of environmental policy and economic costs are key to ensuring that insights used for decision-making are explainable and transparent for environmental scientists and policy-makers (e.g., ~\citep{Sims:2019wg}).

\paragraph{Insurance.} Data science technologies enable institutions to leverage a range of data, including insurance claims, membership and provider data, benefits and medical records, and customer and case data. This data is gathered (preparation), structured (management), processed (modelling and analysis), and turned into valuable insights (dissemination) for the healthcare insurance business. As a result, cost reduction, quality of care, fraud detection and prevention, and consumer engagement can be substantially improved. \cite{Li:2019vu} discuss how the data-centric approach can be used for pension planning while \cite{Diao:2021up} discuss a machine learning method for human mortality prediction, which is a fundamental problem for life insurance; similar approaches would also apply to healthcare insurance.

Accurate claim prediction using data analysis gives insurance companies the ability to minimize financial loss~\citep{Nian:2016tr,Diao:2019tq}. These analysis tools enable the detection of relations between claims and missing observations in order to develop an individual customer’s portfolio. Forecasting upcoming claims allows companies to charge competitive premiums that are not too high and not too low, yielding optimal pricing models for companies and customers.

\paragraph{FinTech.} FinTech refers to modern technology and platforms for financial services. Developments in data science have driven the fast evolution of FinTech. Data science techniques are employed for the efficient analysis of massive financial datasets, including the optimization and design of robo-advisors (explainability), algorithms for blockchain, computational strategies and security of cryptocurrency, and dynamic evaluation of financial services platforms.

\section{Social and Policy Context}
\label{sec:soc}

\improvement[inline]{\normalsize This is incomplete and requires more care. Treat it as a placeholder for now.}

As noted earlier, data science deployments are highly sensitive to the societal and policy contexts in which they are deployed. For example, what can be done with data differs in different jurisdictions. The context can be legal, establishing legal norms for data science deployments, or they can be societal in identifying what is socially acceptable. Furthermore, there are significant intersections between social science and humanities and the core issues in data science. There are four central concerns: ownership, representation, regulation, and public policy. Obviously, there is overlap between these and the data ethics concerns discussed previously.

\paragraph{Ownership.} Data ownership, access, and use -- particularly in terms of how individual data is generated, who owns it, who has access to it, and, by extension, who profits from it -- is a critical concern. At the societal and organizational levels, researchers analyze how economic systems are increasingly data-dependent in terms of both operations and revenue streams, and how pressures to collect and share ever-more intimate data may conflict with users’ own calls for privacy and autonomy. Data privacy from a technical perspective was discussed previously, but it obviously has a significant legal and social dimension that requires careful study (see, e.g.,~\citep{Solove:2006tc}).

\paragraph{Representation.} A primary concern in the development of data science technologies is ensuring diverse and equitable representation at all stages of the lifecycle. This includes the evaluation of the training, tools, and techniques used in data science, including who designs them, who has access to them, and who is represented by them. Data representativeness is tightly tied to questions of marginalization and bias that appear throughout the design, data collection, analysis, and implementation of these technologies -- see, for example, ~\citep{Richardson:2019tf,Richardson:2021aa}.  Another concern is how data is increasingly used to ``speak'' for users, often without their knowledge, changing individuals' relationships with their local communities, corporations, and state. 

\paragraph{Regulation and Accountability Practices.} Components of ethical data science also include a commitment to transparency and explainability in terms of analyzing the inputs of data-driven decision-making, the algorithms applied, and how they lead to specific outputs and recommendations. Ensuring that the benefits and opportunities afforded by advances in data science equally benefit broader society involves accountability and regulatory practices at each stage of the data science lifecycle, and are not only centred on laws and policy interventions, such as the General Data Protection Regulation (GDPR) and the Canadian Personal Information Protection and Electronic Documents Act (PIPEDA). They must also include efforts to include values-in-design, interventions for more accessible and inclusive design, and tools for ethical thought at the levels of training, education, and ongoing daily practice.

\paragraph{Public Policy.} There is a critical and urgent need for the integration of data science in the analysis of public policy~\citep{Steif:2021um,Ozsu:2023ab}. In an age where every Facebook, X (Twitter), and Instagram post is a data observation that can be archived, and can become a part of a historical dataset that can inform public policy, governments have been left behind in their ability to collect, aggregate and analyze this data. With the necessary tools, this data could be managed and analyzed in a way that is explainable and that can be disseminated in a meaningful manner to provide key insights. In a similar vein, there is a paradox in the large amounts of ``open'' data that remains unused, along with concerns of a data deficit in terms of more information that could and should be collected to inform public policy. 

In addition to these concerns, the study of human culture and activities has been reinvigorated by the growth of available data and digital tools. Cultural information, either that has been digitized from earlier material (e.g., a 19th century newspaper that has been scanned, indexed, and made accessible) or that begins its life as a digital object (e.g., social media streams or statistical information), is increasingly accessible to humanists and social scientists~\citep{Milligan:2019tu,Romein:2020vc}. Similarly, scholars now have the ability to present their findings in increasingly innovative ways – interactive visualizations, webpages, databases, 3D environments – and to engage with the broader public in new ways. These trends have given rise to the field known as the digital humanities. A broad term for a wide range of scholarly activity, the \emph{digital humanities} are primarily, but not exclusively, concerned with both understanding how technology is impacting and changing the study of human culture and activity, as well as the changing modes of scholarly communication and impact. Therefore, there is considerable interest in the use of data science techniques, tools and methodologies in studying social science questions~\citep{Foster:2021uf}. 

\section{Data Science lifecycle}
\label{sec:lifecycle}

\begin{figure}[ht]
	\centering
	\includegraphics[width=0.8\linewidth]{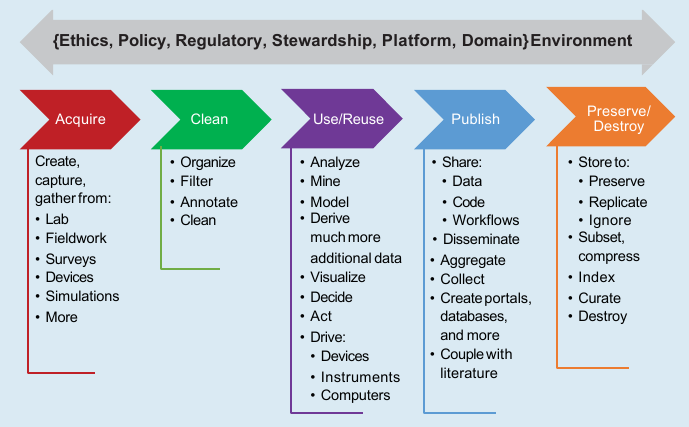}
	\caption{Data lifecycle (Permission to be obtained.)}
	\label{fig:datalife}
\end{figure}

The definition of data science given in Section \ref{sec:ds-def} clearly identifies the process view of data science, namely that it consists of several processing stages starting from data ingestion and eventually leads to improved decisions, insights, and actions. That process is called the \emph{data lifecycle}. The ability to extract insight from massive datasets depends on a comprehensive, systematic, and end-to-end study of the data science lifecycle. The literature refers to ``data lifecycle'', focusing only on data processing. A good definition of data lifecycle is given by the U.S. National Science Foundation Working Group on the Emergence of Data Science~\citep{Berman:2018vj}, which identifies five stages (Figure \ref{fig:datalife}): \emph{acquiring} the data, \emph{cleaning} it and preparing it for analysis, \emph{using} the data through analysis, \emph{publishing} the data and the methods used to analyze the data, and \emph{preserving/destroying} the data according to policy. It recognizes that all of these steps are conducted within the framework of certain ethical norms as well as regulatory and policy framework. This is a relatively general purpose characterization of what happens to data during processing -- not necessarily only in data science applications and deployments.

Variations of this data lifecycle model emerge in various proposals, some predating this formulation. An early proposal, focusing on big data processing, can be found in a white paper prepared for the Computing Community Consortium (CCC). It defines five linear stages~\citep{Agrawal:2012aa}, comprising what is called the ``big data analysis pipeline'': acquisition/recording, extraction/cleaning/annotation, integration/aggregation/representation/ analysis/modeling, and interpretation. The focus is more on the management of the data than on other activities comprising data science as a consequence of the report's focus on big data, which is narrower than data science. ~\cite{Jagadish:2015ul} enhances this lifecycle model by adding feedback loops at each stage and from the final interpretation step to the initial data acquision. Stedman~\citep{Stedman:2021tj} defines a multi-layer model that involves many processing steps at each layer.

Another extension to big data processing has been proposed by~\cite{Erl:2016uq}. Their lifecycle model consists of nine stages: business case evaluation, data identification, data acquisition and filtering, data extraction, data validation and cleansing, data aggregation and representation, data analysis (where iteration is possible), data visualization, and utilization of analysis results. The stages are self-explanatory and disentangles activities that are grouped together in the other models. It provides for iteration at the data analysis stage and between data visualization and data analysis; otherwise it is a linear process from business case to usage.

Stodden~\citep{Stodden:2020wt} extends the data lifecycle by considering, in addition to what happens to the dataset, how that happens (methods, code) and on what type of platform (computational environment). This is an expansive understanding of lifecycle that combines architectural concerns with lifecycle steps and activities. In the current paper, these two concerns are separated and architectural issues are considered in depth in the following section where Stodden's model is discussed further (see Figure \ref{fig:stodden}).

These lifecycle models are very helpful in identifying the steps of the process. However, they give the impression that the entire process is linear and unidirectional. Real project development hardly works in a linear fashion, and Jagadish's revision of the CCC model recognizes that~\citep{Jagadish:2015ul}. An alternative model that is more iterative with built-in feedback loops has been proposed in the CRoss-Industry Standard Process for Data Mining (CRISP-DM) model for data mining projects~\citep{Shearer:2000tm}. It was developed in the second half of 1990s by four companies (Daimler-Benz, Integral Solutions (ICL), which was purchased by SPSS Inc, NCR, and OHRA, which is a large Dutch insurance company) in consultation with 200 users, tool and service providers. CRISP-DM places data at the center and specifies a cyclical life cycle that is iterative and may be repeated over the life time of the project. It  starts with a business problem definition -- what does the business need to drive the mining project. This is followed by data understanding that involves deciding what relevant data is needed and available, and once the relevant data is gathered, it is prepared as discussed in Section \ref{sec:de}. The next step is to select the appropriate analysis model for the task in hand, as discussed in Section \ref{sec:da}, and the analysis is performed. The results are then evaluated to determine the insights and decide on the next steps. The final step is deployment of the analysis model and the insights. The model has feedback loops that allow returning to a previous step until the results are satisfactory. 

\begin{figure}[ht]
	\centering
	\includegraphics[width=0.6\linewidth]{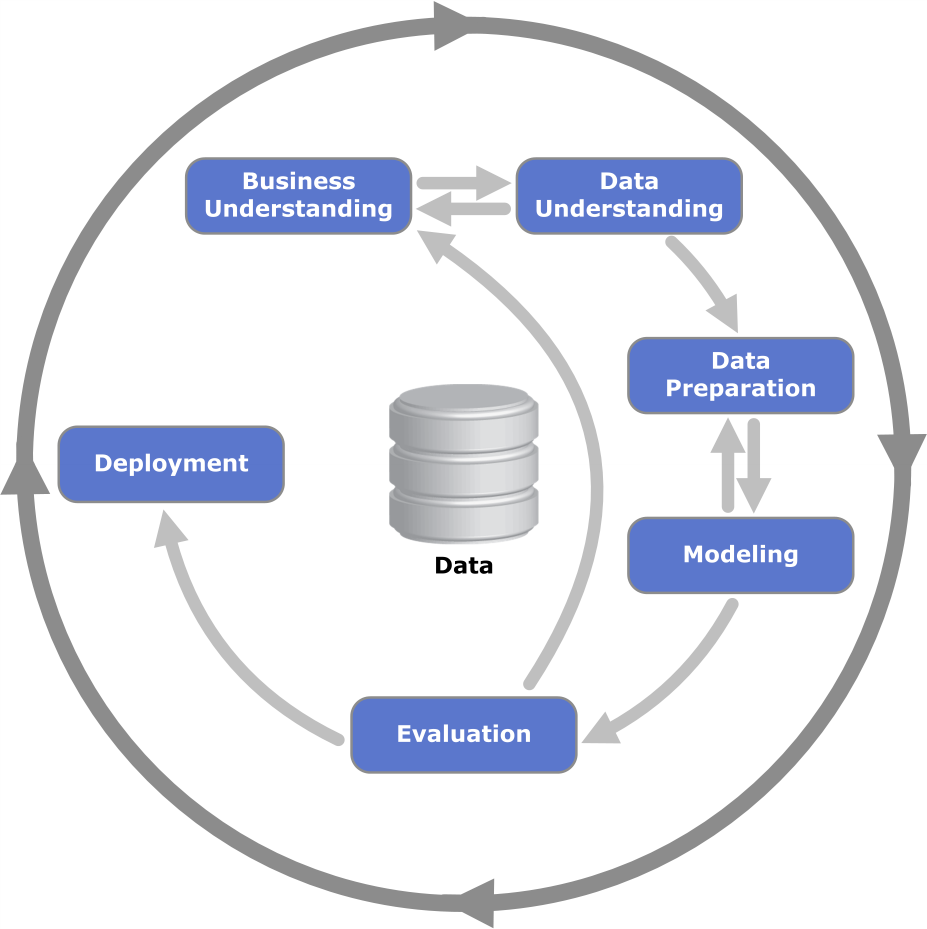}
	\caption{CRISP-DM lifecycle Model (Source: Wikipedia)}
	\label{fig:crisp}
\end{figure}

Over the intervening years, quite a number of ``more modern'' interpretations of the CRISP-DM model have emerged~\citep{Martinez-Plumed:2021}. There are questions as to whether it is still applicable in the big data and data science environment, but it is safe to say that CRISP-DM is still the common lifecycle model for data mining deployments in industry. 

A model similar to CRISP-DM has been proposed for statistical analysis tasks, called PPDAC~\citep{MacKay:2000wj}. The name comes from the phases of the lifecycle: Problem, Plan, Data, Analysis, and Conclusion. PPDAC is simpler than CRISP-DM in that it is a straightforward cycle from Problem through all the phases back to the Problem stage, recognizing that it might be necessary to iterate through the process to refine the problem definition. It has been recognized as a reasonable lifecycle model for these tasks and has been included in statistics books, e.g., ~\citep{Spiegelhalter:2019aa}.

Microsoft Team Data Science Access Life Cycle ~\citep{ms22} also emphasizes the iterative nature of the process. It defines five major stages that are executed iteratively: business understanding, data acquisition and understanding, deployment, modeling, and user acceptance. Business understanding establishe the important parameters that will be considered and the metrics that will be used to determine the success of the project. It also identifies the relevant data sources. Data acquisition and understanding addresses the data engineering tasks identified in Section \ref{sec:de}. Modeling corresponds to the analysis tasks discussed in Section \ref{sec:da}. Deployment pushes the analysis model together with the analysis pipeline to an actual production environment. The fifth and final step is use acceptance.

Data science lifecycle proposed in this paper derives from these iterative models and follows a few principles. First, the architectural issues are separated from the process view of data science -- lifecycle focuses on the stages of the process and how they are interconnected, leaving architectural concerns as a separate discussion. Second, since data science is more than just data, the lifecycle considers the additional dimensions discussed earlier. Third, it recognizes that despite the importance of  applications in framing the investigation, there are data science projects that can be more exploratory where the problem definition is not application-specified but rather answering the question ``what does data reveal?'' or conducting what-if analysis. Fourth, the process  considers the  societal, ethical and policy implications both in the security/privacy dimension and more broadly.

\begin{figure}[ht]
	\centering
	\includegraphics[width=0.6\linewidth]{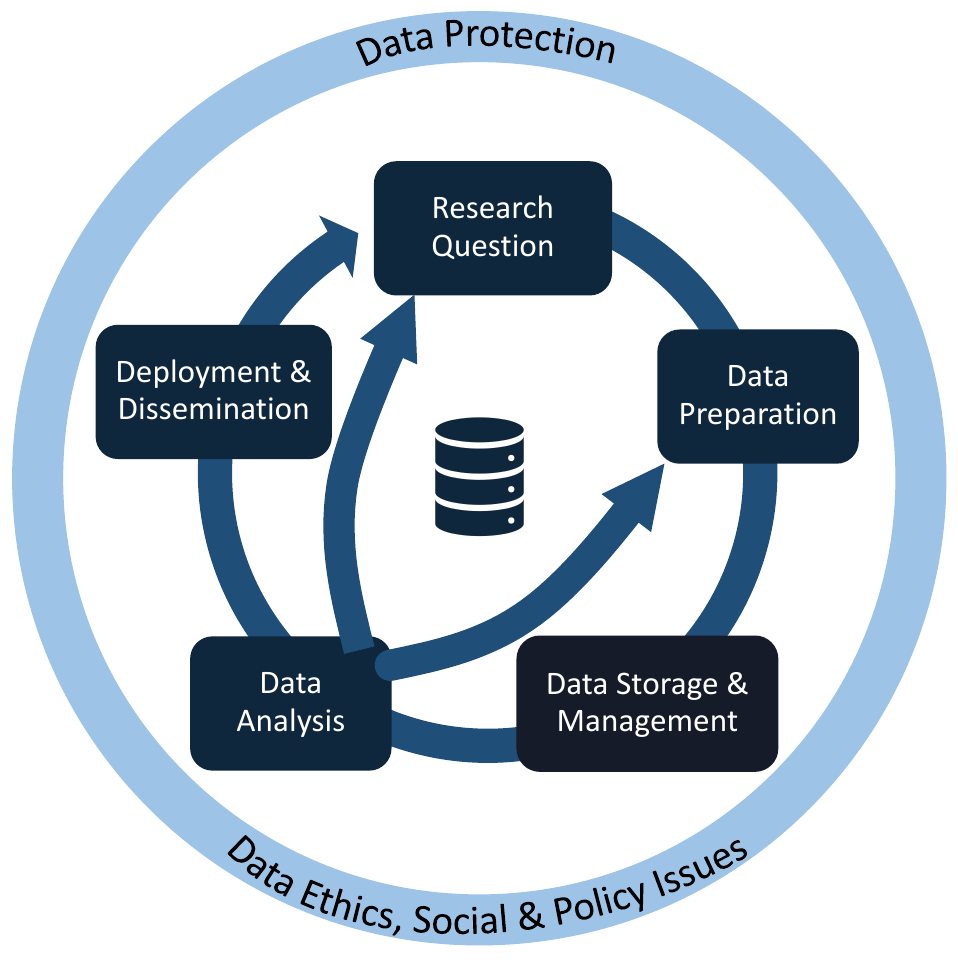}
	\caption{Data Science Lifecycle}
	\label{fig:lifecycle}
\end{figure}

The proposed lifecycle model that follows these principles is shown in Figure \ref{fig:lifecycle}. It starts with the specification of the \emph{research question} that may come from a particular application or may be an exploratory question. A good understanding of the research question is important, since it normally drives the entire process. The next step is \emph{data preparation}, which includes determining which datasets are needed and available, selecting the appropriate datasets from within this larger set; ingesting  the data; addressing data quality issues, including data cleaning and data provenance. The third step is the proper \emph{storage and management} of the data, including the big data management issues discussed in Section \ref{sec:de}. Specifically, data needs to be integrated, decisions need to be made about the storage structures for data for efficient access, appropriate storage structures need to be chosen and designed, suitable access interfaces have to be specified, and provisions need to be made for metadata management, in particular for provenance data. The prepared and suitably stored data is then open for analysis, whose alternatives are discussed in Section \ref{sec:da}. In particular, the appropriate statistical and machine learning model(s) is/are selected/developed, feature engineering is performed to identify the most appropriate model parameters, and model validation studies are conducted to determine the model's suitability. If  model validation is successful, the next step is \emph{deployment and dissemination}, which involves different activities depending on the particular project and application. In some cases, the analysis and processing of data needs to be performed on a continuing basis, so deployment involves maintaining and monitoring the system over time. In other cases, deployment may involve compilation and dissemination of  analysis results and their explanation. Dissemination of the analysis results, and sometimes even the curated data, is an important aspect of this phase. Open data, which is data that ``anyone can freely access, use, modify, and share for any purpose (subject, at most, to requirements that preserve provenance and openness)''\footnote{\url{http://opendefinition.org}} is an important part of dissemination. Many governments and private institutions are adopting Open Data Principles stating that data should not only be open, but should be complete, accurate, primary, and released in a timely manner. These properties make this data very valuable to data scientists, journalists, and  the public. When open data is used effectively, data scientists can explore and analyze public resources, which allows them to question public policy, create new knowledge and services, and discover new (hidden) value useful for social, scientific, or business initiatives.

Problems during analysis phase may result in the process returning to either reformulating the research question (it may be underspecified or over-specified  making model building infeasible) or cycling back to data preparation if the model requires other or different data that has not been prepared. As noted earlier, data science deployments are not ``one-and-done''; following deployment there needs to be constant monitoring -- perhaps the environment changes or the data changes (in the case of streaming data, for example, it is not unusual to observe concept drift that may require revisions to the model), or there is a deeper understanding of the research question that results in its revision and improvement. Thus, the process cycles as a dialectic process -- every time the process comes back to research question, we are at an elevated understanding of what needs to be studied. It is important to recognize that the stages in the lifecycle are not isolated; the boundaries between stages are fuzzy, and there are important and interesting issues that arise at their intersections.

There is a continuous bi-directional interaction between this lifecycle and the data protection issues and the ethics, social and policy concerns. Consider, as an example, data protection concerns and what types of questions they pose on each phase of the lifecycle. For data preparation, the questions include ``Are the current sources willing to provide the data or are there privacy concerns in sharing of the data?'' ``Are there concerns regarding data collectors abusing data?'' For data storage and management the questions might be ``Can data be hidden in a data management system?'' If aggregation is performed when data is stored, ``Do the aggregates reveal things that may be problematic?'' For data analysis, issues such as ``Does the model reveal information about the raw data?'', ``Does the model behave as intended?'', ``How can one control the use of a model?'' or ``How can one obtain a prediction without losing control of its sample?''  need to be considered. At deployment, inferences are being made from the analysis, so questions such as ``How to secure infrastructure?'' and ``How to release data without information leak?'' need to be addressed. 

Similarly, the ethical concerns, and social norms and policy framework have an impact on each of the phases, sometimes even preventing the initiation of particular data science studies. As noted earlier, one of the important characteristics of data science activities is that they always have heavy interaction with some policy and/or societal concern. These projects never occur in a vacuum and are impacted by prevailing policies and societal expectations. These need to be taken into account as they have significant impact. Conversely, data science projects that involve societal questions are likely to have social impact and these need to be considered. Furthermore, these data science activities can  influence policies that are introduced -- what is commonly referred to as data-based or evidence-based decision making.

In this lifecycle model, the focus has been on issues within each stage; this is also true in the discussion of the data science ecosystem (Section \ref{sec:eco}) where the discussion centred around the main issues in each pillar of the ecosystem. This may give the impression that these issues are siloed and can be considered in isolation. That would be wrong and is not what is intended; this organization of the discussion makes it easier to highlight issues, but in reality there is a considerable interplay among the lifecycle stages, and the same holds among the four pillars of the ecosystem. In many cases, the most exciting and challenging problems occur at these interfaces, as the following examples demonstrate. Data visualization may be considered as part of deployment in order to make the data and the results more easily accessible, however, visual analytics is an important part of data analytics and the use of visualization that crosses (or combines) both uses is an important and interesting problem. The interplay between machine learning and data management has become a very popular and is currently heavily studied. The interplay is in both directions -- how can the lessons of data management systems, such as scalability, declarative access, background optimization, may be helpful in building scalable, more functional and more usable machine learning systems? and how can machine learning approaches provide assistance in resolving some thorny data management issues, particularly related to optimization? Of course, there is the long-standing concern regarding the separation of data management and data analysis stages (and systems) that results in data being exported from one and imported into the other every time an analysis needs to be performed. Can this be avoided by pushing analytics inside the data stores and doing this safely? Another point of common overlap has has been repeatedly raised in this paper is between data protection and data ethics.

\section{Data Science System Architecture}
\label{sec:arch}

How should a computing \emph{system} that supports data science activities be structured?  Data science \textit{system} architecture focuses on the data engineering and analysis tasks interwoven with data protection technologies, and in literature they are typically referred to as ``big data architectures.''   

A useful way to think of systems architecture is to first consider a reference architecture rather than a concrete one. A reference architecture is an abstraction that focuses on the functionality and interfaces that should be provided rather than how the system is built (i.e., the concrete architecture); in a sense, it is a template. There can be different concrete architectures that realize this template. In that sense, a reference architecture is useful for discussing system issues at some level of abstraction. It is also helpful to separate specific vendor implementations from the conceptual framework.

\begin{figure}
\centering
\includegraphics[width=0.8\linewidth]{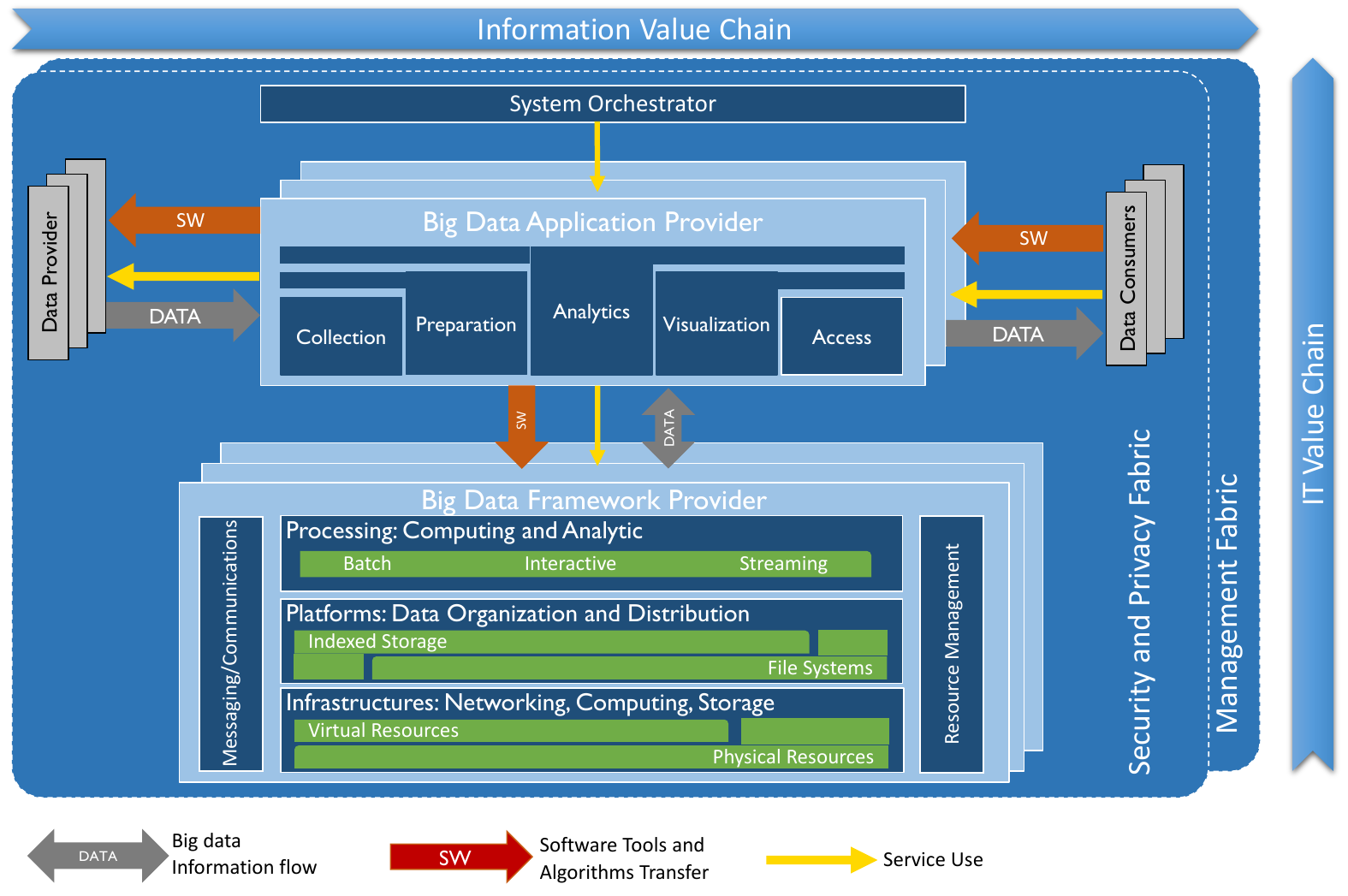}
\caption{NIST NBDRA Reference Architecture}
\label{fig:nbdra}
\end{figure}

A data science system reference architecture is defined by U.S. National Institute of Standards and Technology as part of its Big Data Public Working Group effort to define a big data interoperability framework called NIST Big Data Reference Architecture (NBDRA)~\citep{NIST:2019vf}. This is a comprehensive architectural framework (Figure \ref{fig:nbdra}) that includes lifecycle stages. The NBDRA identifies five main components: big data application provider, big data framework provider, data provider, data consumer, and system orchestrator. As is the case with many reference architectures, these are \textit{logical} components and should be thought of as \textit{roles} that a system should support without any firm commitment as to how these roles are realized. These components are connected by interoperability interfaces and organized into two ``fabrics'' that constitute the dimensions along which the components are organized. The architecture identifies three flows: \emph{data flow}, which can be transmitted either physically or by reference to its location from which it can be obtained; \emph{software flow} that identifies the transmission of software tools so that data can be processed in its original location (i.e., shipping function to data), and \emph{service flow} to represent software programmable interfaces.

A Data Provider is the source of data feed into the system. Since one of the characteristics of big data is the variety, there usually are multiple data providers with different data supplies. These providers can be public data sources or proprietary data holders or aggregators. Data can either flow in raw format, or it can first be processed at the source (e.g., aggregated) and then flow to the target. The interfaces to data providers can vary in their complexity and can implement both push and pull mechanisms.

Big Data Application Provider is responsible for processing the data from its ingest to its conclusion. It executes the lifecycle phases discussed in the previous section. NBDRA defines five stages in the lifecycle: collection, preparation, analytics, visualization, and access, and deals with interactions between these stages by extending some of their activities over multiple stages. For example, data preparation extends over to collection to indicate that some data preparation may take place during data collection; similarly  visualization can be a stage in its own right or be part of providing easy-to-understand access to data and results. These are issues that are already discussed, in greater depth, in the previous section.

The output of Big Data Application Processor (i.e., the processed data and results) are fed to the Data Consumer, which can either be an actual end user or another system. Data Consumer role is the mirror image of the Data Provider in the reverse direction.

The activities from Data Provider, through Big Data Application Provider to Data Consumer are positioned with respect to information value. This is to highlight the fact that value is generated as data is ingested and processed to produce a better understanding or a better outcome.

The fourth component (role) is that of Big Data Framework Provider, which provides the computing and processing platform for big data applications. It identifies three main sub-components organized hierarchically. Infrastructure Framework that focuses on the management of both physical and virtual resources such as networking, computing and storage systems. Above this are the Platforms where the focus is on the software platform, particularly in storing and accessing big data. At the top is Processing that concerns the computing and analytical processing framework, ranging from batch to streaming. Cross-cutting the three layers are Resource Management to manage these complicated computing environments, in particular in scale-out systems, and Messaging that focuses on the communication between different components.

The fifth and final component of the NBDRA framework is System Orchestrator with the responsibility to configure and organize all of these components to obtain a performant, functional, vertical system and to manage it. As noted in the NBDRA document, ``The function of the System Orchestrator is to configure and manage the other components of the Big Data
architecture to implement one or more workloads that the architecture is designed to execute.''

The components from Big Data Framework Provider to System Orchestrator are positioned along the second big data value chain: IT value. This is what is colloquially referred as ``IT food chain'' with the implication that as one moves up the chain, the value of the overall system increases.

As depicted in Figure \ref{fig:nbdra}, NBDRA accommodates multiple occurrences of various components. It is possible to have multiple Big Data Framework Providers and multiple Big Data Application Providers. It also recognizes that every instance does not have to provide all of the sub-components; it is possible to have, for example, one framework provider that provides the infrastructure sub-component and another one that provides the platform sub-component. It is even possible to have two framework providers that implement different platforms. All of this needs to be stitched together into a fully functioning system. Most current realizations of the NBDRA framework are indeed componentized in this fashion. 

NBDRA organizes all of these components within the framework of two ``fabrics'', which are meant to identify the domains of concern. The fabrics identify concerns that cut across all of the components and affect all of them. They are interwoven and interact with each other.

NBDRA is a general and comprehensive reference architecture. It has influenced the development of architectural models that are specialized for given application domains. Klein et al.~\citep{Klein:2016uh} adapt it to develop a reference architecture for applications in the national security domain. That domain-specific focus has resulted in simplifications in certain places and addition of additional fabrics. 

\begin{figure}[ht]
	\centering
	\includegraphics[width=0.8\linewidth]{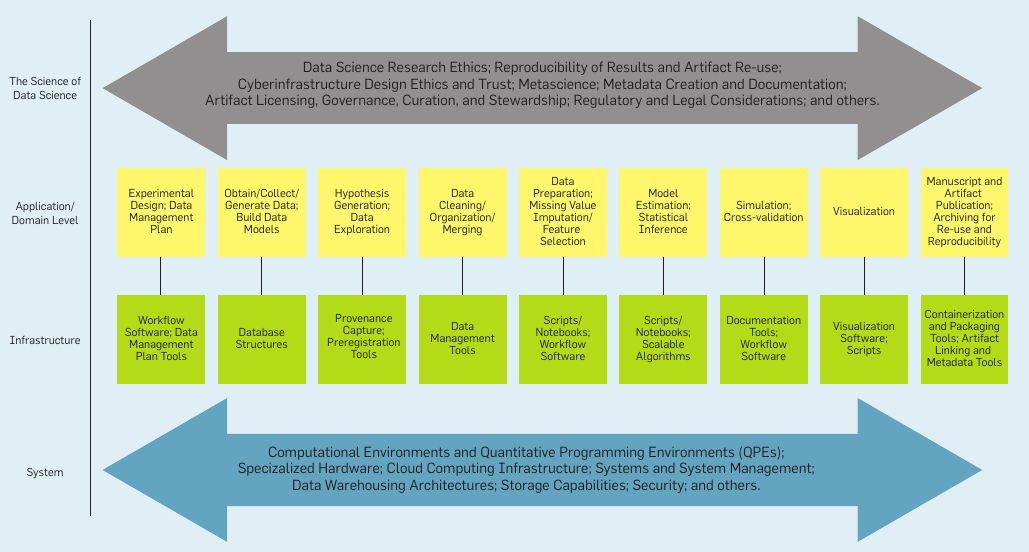}
	\caption{Multi-level Data Science Lifecycle (Permission to be obtained.)}
	\label{fig:stodden}
\end{figure}

\cite{Stodden:2020wt} introduces a model that combines lifecycle issues with architectural concerns, similar to NBDRA. This model identifies four layers (Figure \ref{fig:stodden}) of concern\footnote{Architecture proposals in literature sometimes use the same term to mean somewhat different things. The original terminology is maintained here for fidelity to the proposal. Therefore, it is advisable not to compare terms across different proposals unless explicitly stated.}: System, which identifies the computing environment that is deployed for the project, Infrastructure, which captures the basic computational infrastructure (focusing on software) that is used, Application/Domain, which focuses on the activities that are performed, and the Science of Data Science, where meta-concerns are considered. The System and Infrastructure layers roughly correspond to Big Data Framework Provider in NBDRA, while Application/Domain layer roughly corresponds to Big Data Application Provider. There is no component in NBDRA that corresponds to Stodden's Science of Data Science level, because these concerns are unique to data science and NBDRA focuses on the \textit{system} architecture more than the overall understanding of the ecosystem.

These are not the only architectural models that have been proposed in literature. Avci et al.~\citep{Salma:2017ui,Avci:2020vd} perform domain analysis involving numerous publications that propose architectures. These are consolidated into a reference model that identifies six top-level features: data, data storage, information management, data processing, data analysis, interface, and visualization. Each of these are broken down into more features, each of which form a module. Feature interaction rules  are defined that specify how modules can be combined. A similar literature survey of architectures is by Ataei and Litchfield~\citep{Ataei:2020wc}. P{\"a}{\"a}kk{\"o}nen and Pakkala ~\citep{Paakkonen:2015vx} define a technology-independent reference architecture for big data systems based on analysis of the big data use cases. Many companies have also defined their own architectures that have some of the characteristics of the reference architectures discussed above.

\begin{figure}[ht]
	\centering
	\includegraphics[width=0.8\linewidth]{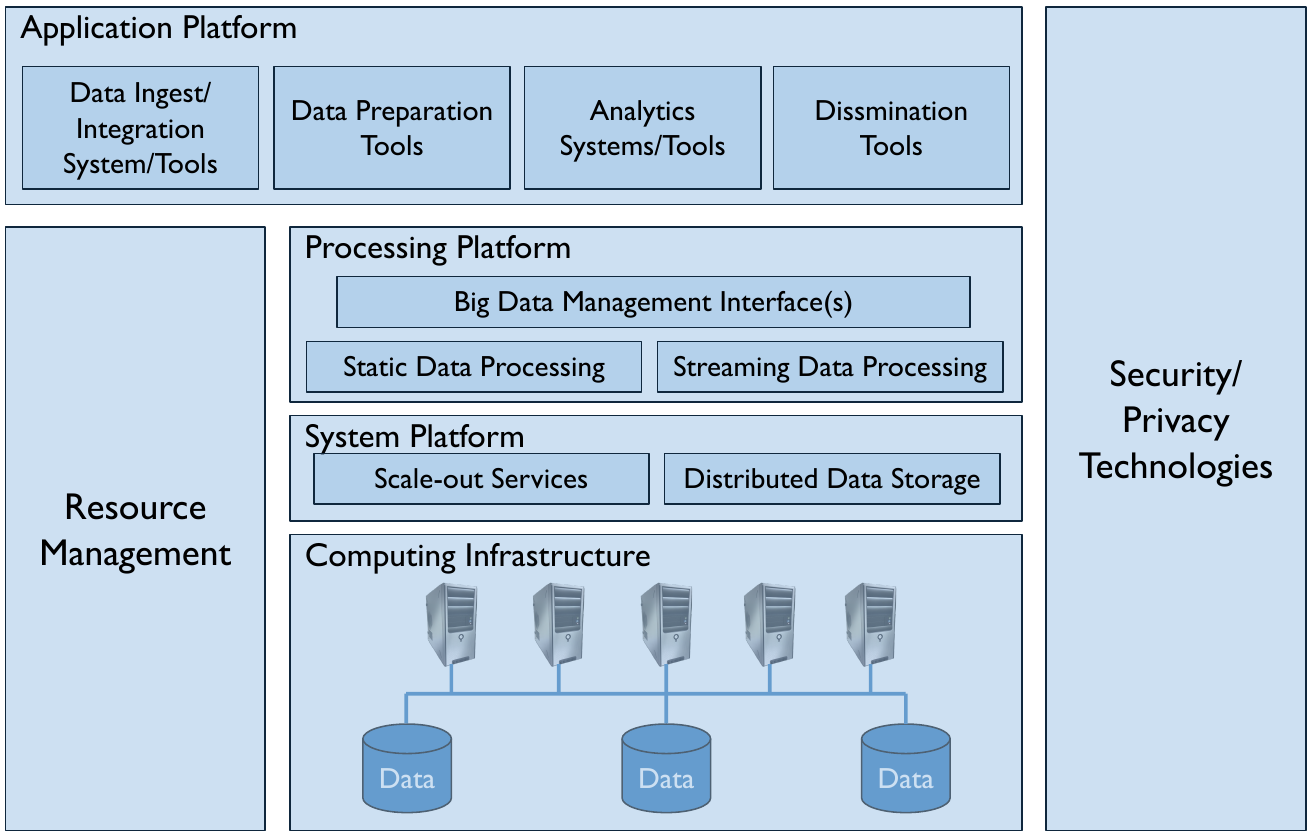}
	\caption{A Possible Concrete System Architecture}
	\label{fig:arch} 
\end{figure}

The reference models are very useful for framing system functionality and roles. However, at some point they need to be mapped to a concrete architecture -- how can a system be actually built that conforms to these references. This concrete architecture identifies specific elements of hardware and software abstractions. The NIST Report discusses possible deployments, but the objective here is to be more concrete. A possible system architecture is depicted in Figure \ref{fig:arch}, identifying hardware and software building blocks. The current trend in building such a system is to have specialized systems and tools for different components and then stitch them together to obtain full functionality.

The  Computing Infrastructure layer of the proposed concrete architecture concerns  the underlying computing systems. Given the size of the data and the heavy computation requirements, these are typically scale-out solutions. Furthermore, they generally involve hardware assist (GPUs, FPGAs, etc). So, the computing infrastructure is typically  heterogeneous~\citep{Zahran:2019ut} organized in a variety of ways: cluster computing, cloud deployment (see below), and even disaggregated architectures that separate processors, storage devices and  memory~\citep{Blagodurov:2021vr}, possibly requiring support for distributed shared memory management~\citep{vldb23_Wang:2022uc}.

System Platform layer includes the software platform that executes on this infrastructure. Two broad categories can be identified: Distributed Data Storage System  (DSS) and Scale-out Services (SoS). DSS provides solutions to store and manage data distributed over multiple storage servers. Two complementary approaches are usually provided: storage as objects or storage as files, with the possibility of hybrids. Object storage manages data as objects. An object includes its data along with a variable amount of metadata, and a unique identifier in a flat object space. Object stores are particularly useful to store a very high number of relatively small data objects, such as photos, mail attachments, etc. Therefore, this approach has been popular with most cloud providers who serve these applications. An important example is Amazon S3 storage system. File storage manages data within unstructured files (i.e., sequences of bytes) on top of which data can be organized as fixed-length or variable-length records. A file system organizes files in a directory hierarchy, and maintains for each file its metadata (file name, folder position, owner, length of the content, creation time, last update time, access permissions, etc.), separate from the content of the file. Thus, the file metadata must first be read to locate the file's content. File storage is appropriate for sharing files locally (e.g., within a data center) and when the number of files are limited (e.g. in the hundreds of thousands). Bigger files that may contain high numbers of records are split (sharded) and distributed on multiple storage nodes requiring a distributed file system. Google File System and its open-source implementation HDFS, are important examples of distributed file systems.

SoS constitutes the second set of services at this layer. These include different services to run applications on a scale-out computing platform, including on cloud platforms. Some example services are task schedulers and monitors (e.g., Kafka Scheduler), data replication solutions (e.g., Microsoft DFS, Opentext Carbonite, Syniti), and in case of cloud deployment, virtual machine management (e.g., Oracle Virtualbox, VMWare, Microsoft Hyper-V) and, perhaps, serverless computing support~\citep{Castro:2019vp}. The exact composition of these services are dependent on the underlying system architecture. Currently, these services and tools are usually merged into systems that operate at the Processing Platform layer, but that is an example of constructing operational systems by vertically integrating software at multiple levels. From the viewpoint of a software stack, it is useful to see them on their own.  

Processing Platform deals with the computational platform for big data processing (see Chapters 11 and 12 in \citep{OzsuV:2020aa} for detailed discussion of this layer). The data is multi-modal, but falls into two categories for processing as discussed previously: static (persistent) data and streaming data.  The data management issues for the former type of data are handled by relational database management systems (for structured data) or by NoSQL systems (for others). In big data processing, NoSQL has been an important complement to relational systems, and this class consists of key-value stores (e.g., DynamoDB), document stores (e.g., MongoDB), wide-column stores (e.g., Google Bigtable), and graph systems (e.g., Neo4j). These run on frameworks such as MapReduce~\citep{li:2013uq} or Spark~\citep{zaharia:2010,zahariabook:2016}. Systems that support streaming data have different requirements -- they deal with data that arrive for processing ordered in time (i.e., a timestamp) and continuously; therefore, data is not stored persistently before processing, but has to be processed as it arrives. Furthermore, since the data arrives continuously, there is no end to it so the processing algorithms cannot wait (block) until they see the entire dataset, they need to be unblocking~\citep{aggarwal:2007qy,golab:2010fj}. These systems typically have real-time processing requirements and are necessary for real-time analytics and decision-making. Current generation systems such as Spark Streaming, Apache Storm, and Apache Flink establish processing platforms for streaming data. These systems can be accessed using their native interfaces or there could be scripting and declarative (SQL-like) querying tools on top of the data processing frameworks.

Managing distributed resources is challenging, and more so in a data science platform with its many complications. The entire stack from Computing Infrastructure to System Platform to Processing Platform require careful management, which is the domain of the Resource Management component. This is also known as ``cluster management'' if the underlying  infrastructure is a computing cluster, and includes tools, methodologies and techniques that enable the scale-out system to operate efficiently and effectively.  Examples of systems that are used for this purpose include Hadoop Yarn, Google Kubernetes and Apache Mesos. In the case of cloud deployment, virtual machine monitoring is an important resource management service as well -- an example being LogicMonitor. 

Above these sits the Application Platform that contains tools and systems that support data processing from ingest to dissemination of the analysis results. The first two sets of functionalities that are provided at this layer (data ingest/integration and data preparation tools and systems) provide the data engineering functionality discussed in detail in Section \ref{sec:de}. All major database system providers (Microsoft, Oracle, IBM and others) are active in this space as well as vendors that specifically target this space such as Tamr, Talend, Trifecta, and Altair. Although these two functionalities are separated, and there are in some cases systems and tools that target only one of them, quite often there is a cross-over of functionalities and the same system may assist in data integration and data preparation. The second component at this layer focuses on data analytics as discussed in Section \ref{sec:da}. There are major systems in this space, such as SAS and Mahoot, but R is also a very popular tool for analysis. Final component of Application Platform are dissemination tools and systems. These include data visualization systems such as Tableau, Microsoft Power BI, and tools that assist in converting data to RDF format (e.g., OpenRefine, R2RML, D2RQ) for creating open data platforms or for feeding into semantic web via Linked Open Data.

As discussed earlier, the security of the end-to-end system and the privacy of the data that is handled is paramount in data science. Therefore, Security/Privacy Technologies span the entire stack addressing the concerns discussed in Section \ref{sec:protect}.

In cloud-based deployments, the layers of this stack can be mapped to specific cloud services. Computing Infrastructure and System Platform together can be deployed on the cloud as an Infrastructure-as-a-Service (IaaS), allowing the creation of virtual machines of appropriate capability for the desired performance level. Processing Platform can be provided as a Platform-as-a-Service (PaaS) enabling a full computing platform with development tools and APIs as a service. Finally the Application Platform can be provided as Software-as-a-Service (SaaS).

\begin{figure}[ht]
	\centering
	\includegraphics[width=0.8\linewidth]{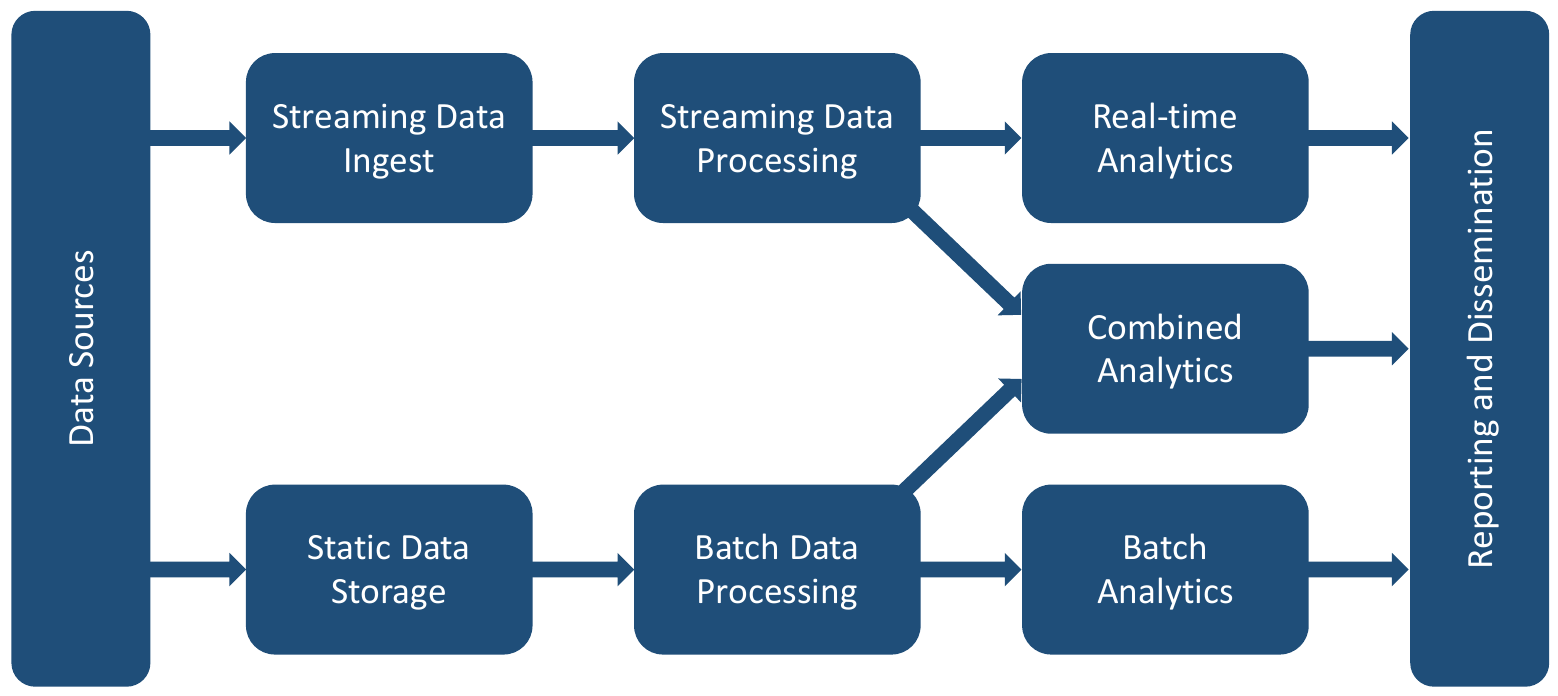}
	\caption{Process-View Architecture}
	\label{fig:process}
\end{figure}

To end this section, it is useful to discuss the process view architecture of data science systems. The process view tracks the data lifecycle discussed in Section \ref{sec:lifecycle}. Figure \ref{fig:process} captures the main characteristics of this perspective. Static data arriving at the system is stored in persistent storage for subsequent processing and undergoes batch processing, which includes both data preparation operations and other normal data access through applications (e.g., database declarative queries, search, or computation). This data subsequently goes through batch analysis to produce results. The streaming data, due to its nature, is not normally stored, but the ingest process does a certain amount of data preparation. The streaming data is also processed, but this is very  different than processing static data: either simple filter and search primitives may be applied as data flows through or they can go more sophisticated processing through windowing. The streaming data is also analyzed and this is called real-time analytics. The streaming data can also be spooled to streaming data warehouses and combined with static data for further analysis that considers both types of data. The results of different analyses are then reported and disseminated.

\section{Data Science is Interdisciplinary}
\label{sec:owner}

Who ``owns'' data science  is a topic of some discussion, primarily between statisticians and computer scientists. This discussion bleeds into the question of who are data scientists, which then leads to different educational models of data science. Given the centrality of data to both disciplines, this discussion is perhaps not surprising. It is worthwhile to address the issus at the end of this article, although the topic of data science education and training is a large one that deserves its separate treatment.

The concern among statisticians regarding the field of data science is long-standing. Given the early promotion of data analytics as an important topic by Tukey (see Section \ref{sec:ds-def}), there is a strong feeling among statisticians that they own (or should own) the topic. In a 2013 opinion piece, ~\cite{Davidian2013} laments the absence of statisticians in a multi-institution data science initiative and asks if data science is not what statisticians do. She indicates that data science is ``described as a blend of computer science, mathematics, data visualization, machine learning, distributed data management--and statistics,'' almost disappointed that these disciplines are involved along with statistics. 

Similarly,  Donoho, in an article entitled \emph{50 Years of Data Science}, laments the current popular interest in data science, indicating that most statisticians view new data science programs as ``cultural appropriation.'' In an extensive treatise, he argues that ``there is a solid case for some entity called `data science' to be created, which would be a true science: facing essential questions of a lasting nature and using scientifically rigorous techniques to attack those questions.'' He continues: ``Insightful statisticians have for at least 50 years been laying the groundwork for constructing that would-be entity as an enlargement of traditional academic statistics. This would-be notion of data science is not the same as the data science being touted today, although there is significant overlap''~\citep{Donoho:2017aa}.

\begin{figure}[H]
\centering
\includegraphics[width=0.5\linewidth]{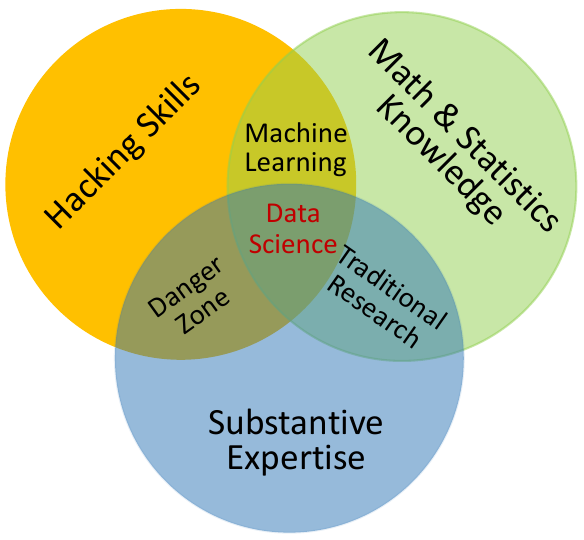}
\caption{Conway View of Data Science}
\label{fig:conway}
\end{figure}

There is a well-known argument put forth by \cite{Conway:2015aa} on the nature of data science. Faced with the difficulty of defining what it is, he instead identifies the knowledge required to engage in the field. He proposes three main areas organized as a Venn diagram (Figure \ref{fig:conway}). The three areas he argues as essential for data science are hacking skills, mathematics and statistics and substantive experience. The hacking skills he argues to be important are the ability ``to manipulate text files at the command-line, understanding vectorized operations, thinking algorithmically.'' Mathematics and statistics knowledge, at the level of ``knowing what an ordinary least squares regression is and how to interpret it'' is required to analyze data. The substantive experience is about the research problem that may come from an application domain or a specific research project. The overlapping areas are interesting: hacking skills plus mathematics and statistics is machine learning, while mathematics and statistics plus substantive experience in an area is the traditional approach to research. The combination of all three areas is data science. He considers hacking skills plus domain experience as the danger area as he identifies those who fall into this area as those ``who know enough to be dangerous.'' The Conway diagram, as it has come to be known, has become popular in these ownership debates by those who don't see a central role for computer science in data science, because Conway argues that hacking skills have nothing to do with computer science: ``This, however, does not require a background in computer science -- in fact, many of the most impressive hackers I have met never took a single CS course.'' 

\begin{figure}[htb]
\centering
\includegraphics[width=0.4\linewidth]{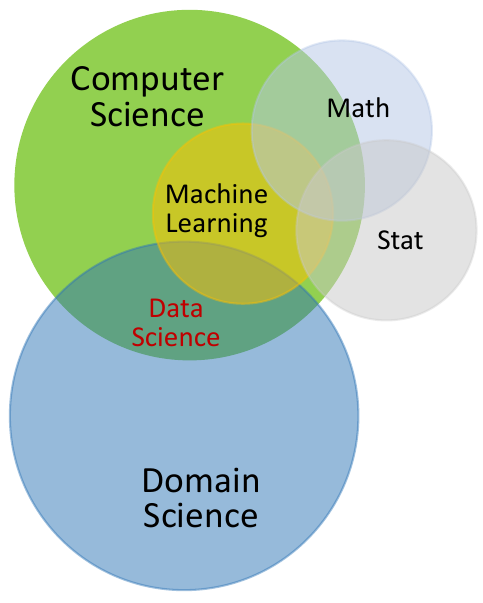}
\caption{Ullman View of Data Science}
\label{fig:ullman}
\end{figure}

Computer science view espouses instead the centrality of computing. One such view has been put forward by \cite{Ullman:2020aa}, who also uses a Venn diagram (Figure \ref{fig:ullman}). He counters Conway, since he considers ``algorithms and techniques for processing large-scale data efficiently as the center of data science.'' Ullman claims that the two big knowledge bases of data science are computer science and domain science (i.e., the application domain), and their intersection is where data science resides. He sees, rightly, machine learning as part of computer science. He argues, again rightly, that some of machine learning is used for data science, but there are applications of machine learning that are outside of data science. His  diagram shows that there are aspects of data science that require computer science techniques that have nothing to do with machine learning -- data engineering as discussed in Section \ref{sec:de} would fall into that category. Some of these points may not be controversial. Where the argument is likely to be challenged is that, in his view, mathematics and statistics ``do not really impact domain sciences directly'' albeit their importance in computer science.

\begin{figure}[htb]
	\centering
	\includegraphics[width=0.4\linewidth]{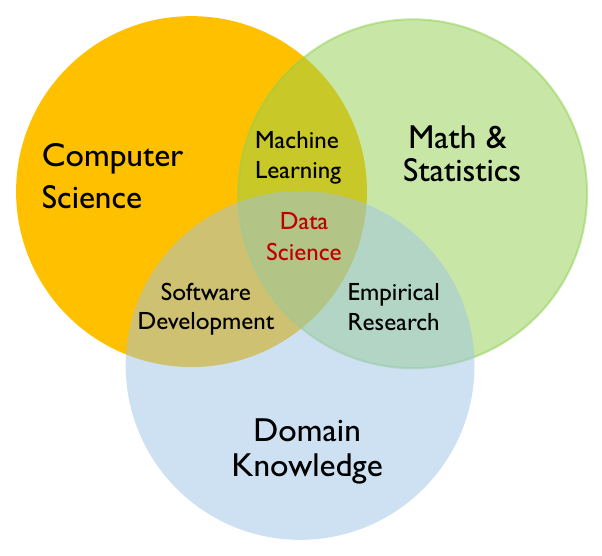}
	\caption{Vogt View of Data Science}
	\label{fig:vogt}
\end{figure}

A more balanced view has been put forth by Marina Vogt\footnote{The original article can no longer be found, but Vogt's viewpoint can be found here: \url{http://www.policyhub.net/node/212}.} that indicates data science as sitting at the intersection of computer science, mathematics and statistics, and domain knowledge (Figure \ref{fig:vogt}). This is more in line with the viewpoint expressed by \cite{Mike:2023aa} and the view put forward by ACM in its curriculum proposal: ``Data science is an interdisciplinary endeavor between computer science, mathematics, statistics, and applied areas such as natural sciences.'' ~\citep{Force:2021aa}. However, even this viewpoint is very STEM-centric and leaves out many topics that are of interest to data science as a field.

\begin{figure}[h]
\centering
\includegraphics[width=0.5\linewidth]{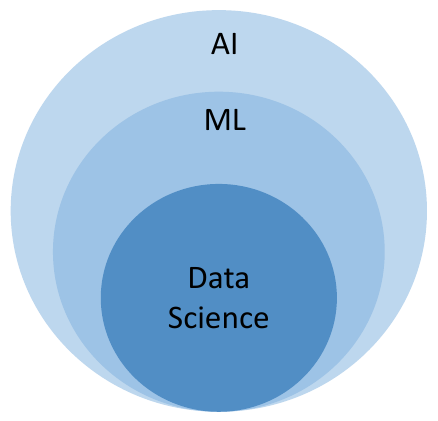}
\caption{AI/ML View of Data Science}
\label{fig:aiml}
\end{figure}

Within computer science, there is a discussion regarding the relationship of AI/ML with data science; this was discussed earlier, but is worth summarizing here for completeness. The argument, primarily by some in the AI/ML community, is that data science is a subset of machine learning, which is a subset of AI (Figure \ref{fig:aiml})\footnote{Venn diagrams seem to be a popular way of arguing these points. There have been a number of follow-ups to Conway's argument, mostly specified as venn diagrams~\citep{Taylor:2017vg}.}. As noted in Section \ref{sec:ds-def}, data science and AI are distinct areas with their own research issues although they intersect when ML techniques are used for analysis and prediction. Outside of this intersection, data science involves a broad set of other research topics that include data quality and trustability, big data management, interpretation and explanation of models and techniques, including visualization, as well as data protection, ethics, and policy. Similarly, AI has a research agenda far broader than ML, despite the significant recent interest in ML.

These discussions and the resulting controversies are not helpful or necessary; they do not move the data science agenda forward. No single community ``owns'' data science -- it is too big, the challenges are too great, and it requires involvement from many disciplines. Most of the ownership discussion is very STEM-centric, but, as argued in this paper, interdisciplinarity of data science extends beyond STEM. Creation of knowledge and use of knowledge is a fundamental human activity that span millenia. This activity is at the core of what we can define as being our collective human culture. Attempts to splinter the core of human achievement through an attribute of ownership is, at its best a narrow parochial view, and at its worst ascendancy of self-interest and greed. 

\begin{figure}
\centering
\includegraphics[width=0.6\linewidth]{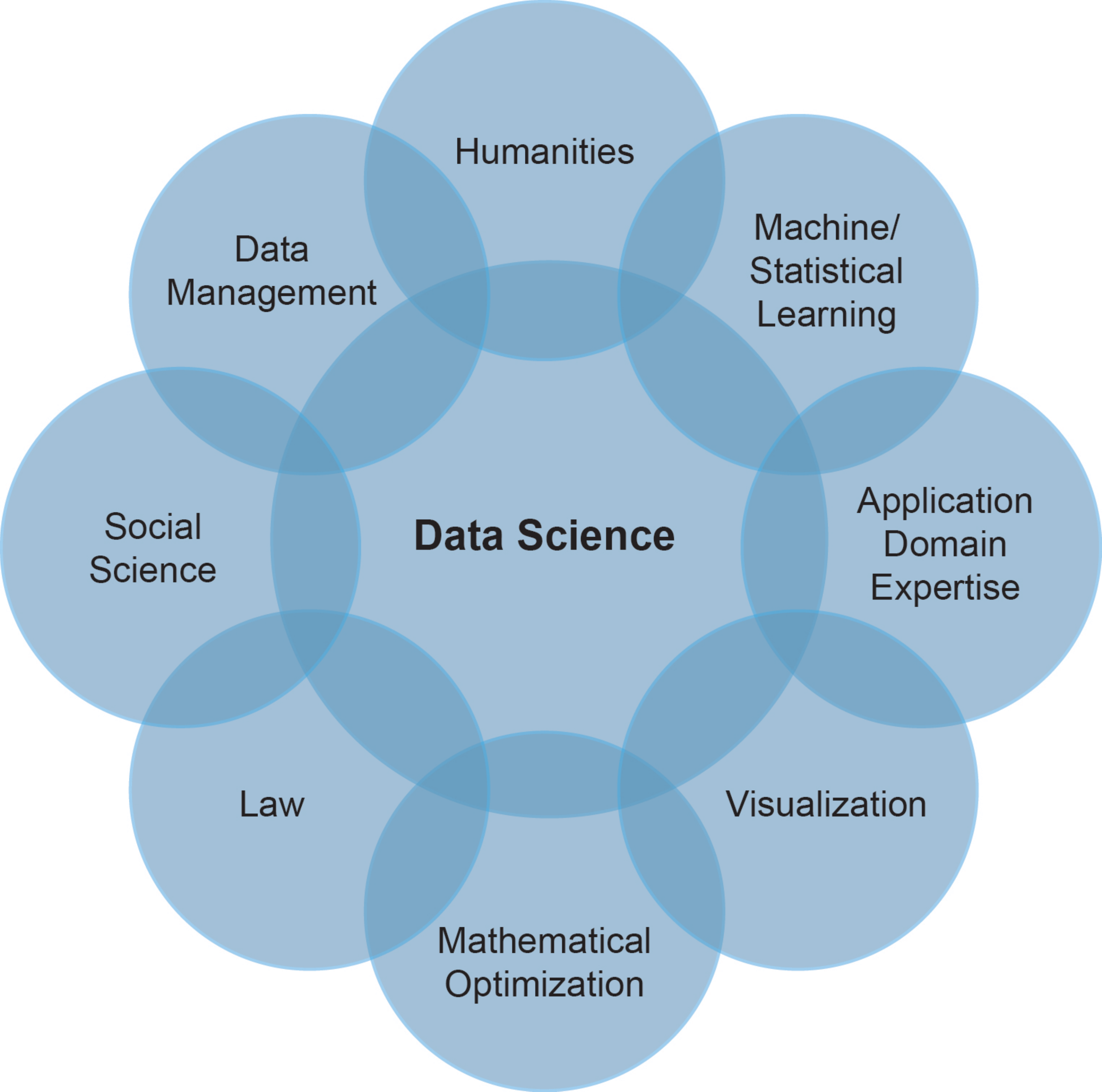}
\caption{Unifying View of Data Science}
\label{fig:unify}
\end{figure}

Data science should be viewed as a unifying force that connects a number of fields (Figure \ref{fig:unify}), some of which are STEM and some not.  I go back to my discussion in Section \ref{sec:intro} of the stakeholders, which are diverse. The ownership arguments take place within one stakeholder group -- STEM people who focus on foundational techniques and the underlying principles. Within this group, it is important to recognize and accept that there are communities with complementary and sometimes overlapping interests: computer scientists who bring expertise in computational techniques/tools that can effectively deal with scale and heterogeneity, statisticians who focus on statistical modeling for analysis, and mathematicians who have much to contribute with discrete and continuous optimization techniques and precise modeling of processes. However, this is only one stakeholder group; there are two others identified in the introduction:  STEM people who focus on science and engineering data science applications, and non-STEM people who focus on social, political, and societal aspects. A danger in such a unifying view is to find the right balance between inclusiveness in accepting the contributions of all these fields and identifying the core of data science. I believe the arguments in earlier sections of this paper have established the core, so this danger is averted.

A natural question that arises from the ownership discussion is ``who is a data scientist?'' The viewpoint differs depending on who the ``owner'' is. The best way to answer this question is to refer to the main stakeholder groups identified above.
When these stakeholders are intersected with the core areas discussed in Section \ref{sec:eco}, a reasonable picture emerges as to what core competencies a data scientist should have:

\begin{itemize}
	\item In-depth knowledge (i.e., expert level) of at least one of the core topics of data engineering or data analytics;
	\item Working knowledge of the other three core topics, sufficient to engage in collaborative work with experts in those topics;
	\item In-depth knowledge of one or more application areas, sufficient to be able to work productively and collaboratively with domain experts;
	\item Ability to work in a team and communicate efficiently.
\end{itemize}

The argument is that a data scientist can be anchored in either data engineering or data analytics. Data protection is not identified as an area where a data scientist can position herself, because data protection concerns cross across all activities and system stack. Similarly, it is not clear how someone can be a data scientist by focusing predominantly on data ethics. This viewpoint, again, establishes a core for the field.

It is important, however, that a data scientist who is anchored in one of those two areas has sufficient knowledge about the other three core building blocks of data science. For example, a data scientist with data engineering focus needs to have sufficient data analytics knowledge to be able to understand data preparation needs for analytics and how different models might impact (and may be impacted) by the data management/data preparation decisions. Conversely, one who is anchored on data analytics needs to understand sufficient data engineering to be able to speak to model requirements in data preparation and understand how the interaction can be most effective. Any data scientist needs to have considerable knowledge of privacy and security concerns and technologies and the governing policies and social norms that are in effect in the particular jurisdiction where the work takes place. Ethics is never out of the picture for a data scientist, who needs to be concerned with both the biases that are implicit in the techniques and data that are used and the broader ethical concerns about deployments.

The third component of this discussion that comes up is how data scientists should be educated and trained. This paper raises this as an issue, but does not engage the topic as it requires a deep and separate treatment of its own. The framing of the education and training programs are heavily influenced by the specification of the core competencies, so the above discussion provides some hint at one possible framing. However, this topic is now the subject of considerable discussion and many institutions are experimenting with different approaches and active scholarly discussions are taking place in publications (e.g., ~\citep{Bonnell:2022uu,icde17_HoweFHKU17,Lau:2022vg,Irizarry:2020uw}). It is best to leave that discussion to a future focused publication.

\section{Conclusions}
\label{sec:conc}
 
Despite its recent popularity, the field of data science is still in its infancy and there is much work that needs to be done to scope and position it properly. The success of early data science applications is evident: from health sciences, where social network analytics have enabled the tracking of epidemics; to financial systems, where guidance of investment decisions are based on the analysis of large volumes of data; to the customer care industry, where advances in speech recognition have led to the development of chatbots for customer service. However, these advances only hint at what is possible; the full impact of data science has yet to be realized. Significant improvements are required in fundamental aspects of data science and in the development of integrated processes that turn data into insight. In particular, current developments tend to be isolated to sub-fields of data science, and do not consider the entire scoping as discussed it in this article. This siloing is significantly impeding large-scale advances. As a result, the capacity for data science applications to incorporate new foundational technologies is lagging.

The objective of this article is to lay out a systematic view of the data science field. To repeat what I said in Section \ref{sec:intro}, key messages are that: (1) it is important to clearly establish a consistent and inclusive view the entire field; (2) in order to avoid becoming a catch-all or whatever the particular circumstances allow, it is essential to define the core of the field while being inclusive; (3) it is critical to  take a holistic view of activities that comprise data science; and (4) a framework needs to be established that facilitates cooperation and collaboration among a number of disciplines. 

\section*{Acknowledgements}
A preliminary version of the ideas in this article  have appeared in an earlier opinion piece in \textit{Quarterly Bulletin of IEEE Technical Committee on Data Engineering} ~\citep{Ozsu:2020aa}. A shorter version of this article has appeared in \textit{Communications of the ACM} ~\citep{Ozsu:2023aa}.

My views on data science were sharpened in discussions with many colleagues as we worked on a number of data science proposals. I would like to acknowledge the many early discussions on framing the field and the relevant joint work with Raymond Ng and Nancy Reid. I thank Samer Al-Kiswani, Angela Bonifati, Khuzaima Daudjee, Maura Grossman, John Hirdes, Florian Kerschbaum, Jatin Matwani, Renée Miller, and Patrick Valduriez  for feedback on the whole or parts of the paper. Many domain scientist colleagues contributed to my understanding of the applications and their connection to data science. They have contributed initial drafts for discussions in Section \ref{sec:apps}, but I summarized and sometimes restructured them (all errors are mine). I acknowledge the assistance of Jatin Matwani (power and energy), Andrew Doxley and Bin Ma (biological and biomedical systems), Joel Dubin (health sciences and health informatics), Ian Milligan (digital humanities), Thomas Coleman, Yuying Li, Chengguo Weng (Finance and Insurance).
%\end{acks}

\clearpage

\bibliographystyle{abbrvnat}

%\balance % balancing ref page upon ACM suggestion

\fontsize{11}{13.2}
\selectfont
\bibliography{ds-references}

\clearpage

\listoftodos

\end{document}